\documentclass[letterpaper,showpacs,twocolumn,eqsecnum,superscriptaddress,floatfix]{revtex4}

\usepackage{amssymb}
\usepackage{amsmath}
\usepackage{latexsym}
\usepackage{epsfig}

\begin{document}

%%%%%%%%%%%%%%%%%%
%%%   MACROS   %%%
%%%%%%%%%%%%%%%%%%

% Some macros for lazy typing.

\newcommand{\tg}{\tilde{\gamma}}
\newcommand{\tG}{\tilde{\Gamma}}
\newcommand{\tA}{\tilde{A}}
\newcommand{\tR}{\tilde{R}}
\newcommand{\tnabla}{\tilde{\nabla}}

\newcommand{\hg}{\hat{\gamma}}
\newcommand{\hG}{\hat{\Gamma}}
\newcommand{\hA}{\hat{A}}
\newcommand{\hR}{\hat{R}}
\newcommand{\hD}{\hat{\Delta}}
\newcommand{\hnabla}{\hat{\nabla}}

\newcommand{\fg}{\mathring{\gamma}}
\newcommand{\fG}{\mathring{\Gamma}}
\newcommand{\fR}{\mathring{R}}
\newcommand{\fnabla}{\mathring{\nabla}}

\newcommand{\lb}{\pounds_\beta}

%%%%%%%%%%%%%%%%%
%%%   TITLE   %%%
%%%%%%%%%%%%%%%%%

\title{Formulations of the 3+1 evolution equations in curvilinear
coordinates}

\author{Miguel Alcubierre}
\email{malcubi@nucleares.unam.mx}

\author{Martha D. Mendez}
\email{marthadonaji.mendez@nucleares.unam.mx}

\affiliation{Instituto de Ciencias Nucleares, Universidad Nacional
Aut\'onoma de M\'exico, A.P. 70-543, M\'exico D.F. 04510, M\'exico.}

%%%%%%%%%%%%%%%%
%%%   DATE   %%%
%%%%%%%%%%%%%%%%

\date{\today}

%%%%%%%%%%%%%%%%%%%%
%%%   ABSTRACT   %%%
%%%%%%%%%%%%%%%%%%%%

\begin{abstract}
Following Brown~\cite{Brown:2009dd}, in this paper we give an overview
of how to modify standard hyperbolic formulations of the 3+1 evolution
equations of General Relativity in such a way that all auxiliary
quantities are true tensors, thus allowing for these formulations to
be used with curvilinear sets of coordinates such as spherical or
cylindrical coordinates.  After considering the general case for both
the Nagy-Ortiz-Reula (NOR) and the Baumgarte-Shapiro-Shibata-Nakamura
(BSSN) formulations, we specialize to the case of spherical symmetry
and also discuss the issue of regularity at the origin.  Finally, we
show some numerical examples of the modified BSSN formulation at work
in spherical symmetry.
\end{abstract}

%%%%%%%%%%%%%%%%
%%%   PACS   %%%
%%%%%%%%%%%%%%%%


%%%%%%%%%%%%%%%%%%%%%%
%%%   MAKE TITLE   %%%
%%%%%%%%%%%%%%%%%%%%%%

\maketitle

%%%%%%%%%%%%%%%%%%%%%%%%
%%%   INTRODUCTION   %%%
%%%%%%%%%%%%%%%%%%%%%%%%

\section{Introduction}
\label{sec:introduction}

In 3+1 formalism of General Relativity one splits spacetime into a
foliation of 3-dimensional (3D) spacelike hypersurfaces (assuming
that the spacetime is globally hyperbolic), and projects the Einstein
field equations in the normal and tangential direction to those.  In
this way, the 10 independent field equations are naturally separated
into 4 constraint equations and 6 evolution equations for the geometric
degrees of freedom.  The evolution equations that are obtained
directly from this projection are known as the Arnowitt-Deser-Misner
(ADM) equations~\cite{Arnowitt62,York79,Alcubierre08a}.  As was
already realized in the late 80's and early 90's, these ADM evolution
equations, though physically correct, have nevertheless one serious
drawback: they turn out to be only weakly hyperbolic and as such are
not mathematically well-posed (see {\em e.g.}~\cite{Alcubierre08a}).
By this one means that the solutions do not depend continuously on the
initial data and can be unstable in the presence of constraint
violations, which in practice implies that one will encounter serious
stability problems in numerical evolutions based on these equations.

It turns out, however, that one can construct alternative formulations
of the evolution equations by adding to them multiples of the
constraints in a variety of different ways.  These new systems of
evolution equations will have the same physical (constraint
satisfying) solutions, but will typically differ significantly in
their mathematical structure.  Over the last two decades, a number of
different well-posed strongly hyperbolic formulations of the 3+1
evolution equations have been proposed, and several of them have been
tested in numerical evolution codes~\footnote{We will not attempt to
  give an exhaustive list here of such hyperbolic formulations since
  there are simply too many of them. But see {\em
    e.g.}~\cite{Alcubierre08a} for an extended list of references.}.
In particular, the formulation proposed by Shibata and Nakamura, and
Baumgarte and Shapiro, known as the BSSN
formulation~\cite{Shibata95,Baumgarte:1998te}, has turned out to be
very stable and robust in practice, and has become the standard
formulation used by most 3+1 evolution codes today. This formulation
has finally allowed the accurate simulation of binary black-hole
systems with different masses and spins, starting from wide
separations through the merger and ring-down of the final black
hole~\cite{Campanelli:2005dd,Campanelli:2006gf,Baker:2005vv} (one
should mention, however, that the first successful simulation of
multiple orbits of binary black holes was in fact carried out by
F. Pretorius using a very different approach based on the so-called
generalized harmonic formulation, an approach that is still being used
today by a number of different groups~\cite{Pretorius:2005gq}).

The BSSN formulation, though very successful in practice, has the
drawback of involving dynamical quantities that are not true tensors,
such as tensor densities and contracted Christoffel symbols.  This
represents no problem in most 3D simulations where one typically uses
Cartesian coordinates, but becomes an important issue when one
considers curvilinear coordinate systems, such as spherical or
cylindrical coordinates.

Recently, Brown introduced a more general version of the BSSN system
where all dynamical quantities are true tensors~\cite{Brown:2009dd}.
This ``generalized BSSN'' formulation is thus ideally suited for the
use of curvilinear systems of coordinates, which in particular allows
one to construct a BSSN version of the evolution equations for the
case of spherical or cylindrical symmetry.

In this paper we give an overview of the main ideas behind Brown's
approach, and apply them to both the Nagy-Ortiz-Reula
(NOR)~\cite{Nagy:2004td} and BSSN formulations.  The paper is
organized as follows.  In Section~\ref{sec:3+1} we give a brief review
of the 3+1 formalism.  Later, in Section~\ref{sec:background} we
discuss some important results related to the fully covariant
expressions of the Riemann and Ricci curvature tensors in terms of a
background metric.  Section~\ref{sec:NOR-general} then considers the
case of the NOR formulation and its generalization to curvilinear
coordinates.  In Section~\ref{sec:BSSN-general} we repeat the same
analysis for the BSSN formulation, and also include a brief discussion
of the Gamma driver shift condition. In Section~\ref{sec:spherical} we
consider the particular case of BSSN in spherical symmetry, and
discuss the basic equations and the important issue of the
regularization at the origin.  Finally, in Section~\ref{sec:examples}
we present some numerical examples.  We conclude in
Section~\ref{sec:conclusions}.

Throughout the paper we will use geometric units such that
$G=c=1$. Also, Greek indices will represent all spacetime dimensions
and will run from 0 to 3, while Latin indices will represent only
spatial dimensions and will run from 1 to 3.

%%%%%%%%%%%%%%%
%%%   ADM   %%%
%%%%%%%%%%%%%%%

\section{Basic 3+1 equations}
\label{sec:3+1}

Before considering the NOR and BSSN formulations it is convenient to
first review the basic concepts and equations of the 3+1 formalism of
general relativity (for a more detailed introduction see {\em
e.g.}~\cite{Alcubierre08a}).

In the 3+1 formulation spacetime is foliated into spatial
hypersurfaces parametrized by a {\em time function}\/ $t$.  The basic
dynamical quantities of the 3+1 formulation are then the metric of the
spatial hypersurfaces $\gamma_{ij}$ and the extrinsic curvature tensor
of those hypersurfaces $K_{ij}$ which is defined as
\begin{equation}
K_{\mu \nu} := - P^\alpha_\mu \: \nabla_\alpha n_\nu 
\end{equation}
where $n^\mu$ is the time-like normal vector to the spatial
hypersurfaces, and \mbox{$P^\alpha_\beta := \delta^\alpha_\beta +
n^\alpha n_\beta$} the projection operator onto the hypersurfaces.

Furthermore, one also introduces the lapse function $\alpha$ that
measures the proper time elapsed between adjacent hypersurfaces along
the normal direction, and the shift vector $\beta^i$ that controls how
the spatial coordinates propagate from one hypersurface to the next.
In more detail, an observer moving along the normal direction to the
hypersurfaces (also known as an {\em Eulerian}\/ observer) will have a
coordinate speed given by $-\beta^i$, and will measure a proper time
$d \tau = \alpha dt$.  In terms of these coordinates, the unit normal
vector becomes \mbox{$n^\mu = ( 1/\alpha, -\beta^i/\alpha)$}, and the
extrinsic curvature tensor takes the form
\begin{equation}
K_{ij} = - \frac{1}{2 \alpha} \left( \partial_t \gamma_{ij}
- \lb \gamma_{ij} \right) \; ,
\end{equation}
with $\lb$ the Lie derivative with respect to the shift vector, and
where we have only considered spatial components using the fact
that the extrinsic curvature is by definition normal to the
hypersurfaces.
 
Given the spacetime foliation just described, the Einstein field
equations separate naturally into two distinct groups.  The first
group corresponds to those equations that have no time derivatives and
results in the so-called Hamiltonian and momentum constraints
\begin{eqnarray}
H &:=& \frac{1}{2} \left( R + K^2 - K_{ij} K^{ij} \right)
- 8 \pi \rho = 0 \; , \label{eq:hamiltonian} \\
M^i &:=& \nabla_j \left( K^{ij} - \gamma^{ij} K \right)
- 8 \pi j^i = 0 \; . \label{eq:momentum}
\end{eqnarray}
In the above equations $R := \gamma^{ij} R_{ij}$ is the trace of the
spatial Ricci tensor $R_{ij}$, \mbox{$K := \gamma^{ij} K_{ij}$} is the
trace of the extrinsic curvature, and $\nabla_i$ is the covariant
derivative associated with the spatial metric $\gamma_{ij}$, while
$\rho$ and $j^i$ are the energy and momentum densities measured by the
Eulerian observers and are given by
\begin{eqnarray}
\rho &:=& n^\mu n^\nu T_{\mu \nu} \\
j^i &:=& - P^{i \mu} n^{\nu} T_{\mu \nu} \; ,
\end{eqnarray}
where $T_{\mu \nu}$ is the stress-energy tensor of the matter.

The second group of field equations corresponds to the true evolution
equations of the system. In terms of the quantities introduced above
these evolution equations take the form
\begin{eqnarray}
\partial_t \gamma_{ij} - \lb \gamma_{ij} &=& - 2 \alpha K_{ij} \; ,
\label{eq:gammadot} \\
\partial_t K_{ij} - \lb K_{ij} &=& - \nabla_i \nabla_j \alpha \nonumber \\
&+& \alpha \left[ R_{ij} + K K_{ij} - 2 K_{ik} K^k_j \right]
\nonumber \\
&+& 4 \pi \alpha \left[ \gamma_{ij} \left( S - \rho \right)
- 2 S_{ij} \right] \; , \;
\label{eq:Kdot}
\end{eqnarray}
where $R_{ij}$ is the 3-dimensional Ricci tensor associated with the
spatial metric $\gamma_{ij}$:
\begin{eqnarray}
R_{ij} &=& - \frac{1}{2} \: \gamma^{mn} \partial_m \partial_n \gamma_{ij}
+ \gamma_{m(i} \partial_{j)} \Gamma^m + \Gamma^m \Gamma_{(ij)m} \nonumber \\
&+& 2 \Gamma^{mn}{}_{(i} \Gamma_{j)mn} + \Gamma_{mni} \Gamma^{mn}{}_j \; ,
\label{eq:Ricci}
\end{eqnarray}
with $\Gamma^i := \gamma^{mn} \Gamma^i{}_{mn}$, and where $S_{ij}$ is
the stress tensor measured by the Eulerian observers defined as
\begin{equation}
S_{ij} := P^\alpha_i P^\beta_j T_{\alpha \beta} \; ,
\end{equation}
with $S := \gamma^{ij} S_{ij}$.  The evolution equations above are
known in the numerical relativity community as the
Arnowitt--Deser--Misner (ADM) equations~\cite{Arnowitt62,York79}.

%%%%%%%%%%%%%%%%%%%%%%%%%%%%%
%%%   BACKGROUND METRIC   %%%
%%%%%%%%%%%%%%%%%%%%%%%%%%%%%

\section{Curvature tensor in terms of a background metric}
\label{sec:background}

It is convenient at this point to review some well-known fully
covariant expressions for the Riemann and Ricci curvature tensors in
terms of a background metric.

Let us assume that we have a manifold with some coordinate system and
two different metric tensors defined on it: the ``physical'' metric
$\gamma_{ij}$, and some ``background'' metric $\fg_{ij}$ that is not
necessarily flat (though in the following sections we will assume that
the background metric is indeed flat).  We now want to express the
curvature tensor associated with the physical metric $\gamma_{ij}$ in
terms of the curvature associated to the background metric $\fg_{ij}$,
together with covariant derivatives of $\gamma_{ij}$ with respect to
this background. In order to do this, we start by defining the
quantity:
\begin{equation}
\Delta^a{}_{bc} := \Gamma^a{}_{bc} - \fG^a{}_{bc} \; ,
\label{eq:Deltadef1}
\end{equation}
with $\Gamma^a{}_{bc}$ and $\fG^a{}_{bc}$ the Christoffel symbols
associated with $\gamma_{ij}$ and $\fg_{ij}$ respectively.  Notice
that even though neither $\Gamma^a{}_{bc}$ nor $\fG^a{}_{bc}$ are
components of tensors, their difference $\Delta^a{}_{bc}$ is in fact a
proper tensor.

Having defined $\Delta^a{}_{bc}$ let us now calculate the covariant
derivative of the physical metric $\gamma_{ij}$ in the background
geometry, that is $\fnabla_a \gamma_{bc}$.  Notice first that, in
general
\begin{eqnarray}
\nabla_a \gamma_{bc} &=& \fnabla_a \fg_{bc} = 0 \; ,
\label{eq:covdiffg} \\
\fnabla_a \gamma_{bc} &\neq& 0 \; .
\end{eqnarray}
If we now take the convention that indices of $\Delta^a{}_{bc}$ are
raised and lowered with the physical metric $\gamma_{ij}$, then we can
use~\eqref{eq:covdiffg} to show that:
\begin{equation}
\fnabla_a \gamma_{bc} = 2 \Delta_{(bc)a} \; ,
\label{eq:flatcovdiffg}
\end{equation}
and equivalently
\begin{equation}
\fnabla_a \gamma^{bc} = - 2 \Delta^{(bc)}{}_a \; .
\label{eq:flatcovdiffgup}
\end{equation}
One can now solve for $\Delta^a{}_{bc}$ from the above expressions to
find
\begin{equation}
\Delta^a{}_{bc} = \frac{1}{2} \: \gamma^{am} \left(
\fnabla_b \gamma_{cm} + \fnabla_c \gamma_{bm} -
\fnabla_m \gamma_{bc}  \right) \; .
\label{eq:Delta-nabla}
\end{equation}
Notice that this expression for $\Delta^a{}_{bc}$ is in fact identical
to that for the Christoffel symbols $\Gamma^a{}_{bc}$, but with the
partial derivatives replaced with covariant derivatives on the
background.  In particular, if the background is flat and we use
Cartesian coordinates we will have \mbox{$\Delta^a{}_{bc} =
  \Gamma^a{}_{bc}$} and \mbox{$\fnabla_a = \partial_a$}, so that the
last expression reduces to the standard definition of the Christoffel
symbols.

We can now use~\eqref{eq:Delta-nabla} to show that the physical Riemann
curvature tensor can be written in terms of the $\Delta^a{}_{bc}$ as:
\begin{equation}
R^a{}_{bcd} = \fR^a{}_{bcd} + 2 \fnabla_{[c} \Delta^a{}_{d]b}
+ 2 \Delta^a{}_{m[c} \Delta^m{}_{d]b} \; ,
\label{eq:Riemann-nabla}
\end{equation}
where $\fR^a{}_{bcd}$ is the curvature tensor of the
background. Again, the second term has the same structure as the
standard expression for the Riemann tensor, but with the
$\Gamma^a{}_{bc}$ replaced with $\Delta^a{}_{bc}$, and the partial
derivatives replaced with covariant derivatives on the background.
The expression again reduces to the usual one for a flat background in
Cartesian coordinates.

Next, let us lower the first index in the Riemann tensor.  This is not
as trivial as it might seem since now $\gamma_{ij}$ can not be brought
inside the operator $\fnabla_a$ directly. After a somewhat lengthy
algebra, where one needs to use the expression for the commutator of
the covariant derivatives of a rank 2 tensor in terms of the Riemann,
one finally finds that:
\begin{eqnarray}
R_{abcd} &=& \frac{1}{2} \left[ \gamma_{am} \fR^m{}_{bcd}
- \gamma_{bm} \fR^m{}_{acd} \right. \nonumber \\
&& + \left. \fnabla_c \fnabla_b \gamma_{ad}
- \fnabla_d \fnabla_b \gamma_{ac} \right. \nonumber \\
&& + \left. \fnabla_d \fnabla_a \gamma_{bc}
- \fnabla_c \fnabla_a \gamma_{bd} \right] \nonumber \\
&& + \left. \Delta_{mad} \Delta^m{}_{bc} -  \Delta_{mac} \Delta^m{}_{bd}
\right. \: .
\label{eq:Riemannlow-nabla}
\end{eqnarray}
Notice that in the last expression we {\em do not}\/ lower the first index
of $\fR^a{}_{bcd}$ with $\gamma_{ij}$, since by convention it should be
lowered with $\fg_{ij}$.

Finally, let us find the expression for the Ricci tensor
\mbox{$R_{ab} := \gamma^{cd}
  R_{acbd}$}. Using~\eqref{eq:Riemannlow-nabla} we find, after some
algebra:
\begin{eqnarray}
R_{ab} &=& - \frac{1}{2} \: \gamma^{mn} \left[ 
  \fnabla_m \fnabla_n \gamma_{ab}
+ \fnabla_a \fnabla_b \gamma_{mn} \right. \nonumber \\
&& - \left. \fnabla_a \fnabla_m \gamma_{bn}
- \fnabla_b \fnabla_m \gamma_{an} \right] \nonumber \\
&& + \left. \Delta_{mna} \Delta^{mn}{}_b -  \Delta_{mab} \Delta^m 
\right. \nonumber \\
&& - \gamma^{mn} \fR^c{}_{mn(a} \gamma_{b)c} \; ,
\label{eq:Ricci-nabla1}
\end{eqnarray}
where we have defined
\begin{equation}
\Delta^m := \gamma^{ab} \Delta^m{}_{ab} = \Gamma^m - \gamma^{ab} \fG^m{}_{ab} \; .
\label{eq:Deltadef2}
\end{equation}
We can in fact rewrite the Ricci tensor in terms of derivatives of the
quantity $\Delta^m$ just defined.  Using~\eqref{eq:Delta-nabla} one
finds that \eqref{eq:Ricci-nabla1} is entirely equivalent to
\begin{eqnarray}
R_{ab} &=& - \frac{1}{2} \: \gamma^{mn} \fnabla_m \fnabla_n \gamma_{ab}
+ \gamma_{m(a} \fnabla_{b)} \Delta^m + \Delta^m \Delta_{(ab)m} \nonumber \\
&+& 2 \Delta^{mn}{}_{(a} \Delta_{b)mn} + {\Delta^{mn}}_a \Delta_{mnb} 
 \nonumber \\
&-& \gamma^{mn} \fR^c{}_{mn(a} \gamma_{b)c} \; ,
\label{eq:Ricci-nabla2}
\end{eqnarray}
For a flat background in Cartesian coordinates, this last expression
clearly reduces to the standard expression given in~\eqref{eq:Ricci}.

%%%%%%%%%%%%%%%%%%%%%%%%
%%%   NOR - GENERAL  %%%
%%%%%%%%%%%%%%%%%%%%%%%%

\section{The NOR formulation}
\label{sec:NOR-general}

\subsection{Standard formulation}
\label{sec:NOR-standard}

The Nagy--Ortiz--Reula (NOR) formulation~\cite{Nagy:2004td} is in
essence a generalization of the Bona--Masso formulation (BM) of the
early 1990s~\cite{Bona89,Bona92,Bona93,Bona94b,Bona97a}.  This
formulation is based on first writing the three-dimensional Ricci
tensor that appears in the ADM evolution equations as:
\begin{eqnarray}
R_{ij} &=& - \frac{1}{2} \: \gamma^{mn} \partial_m \partial_n \gamma_{ij}
+ \gamma_{k(i} \partial_{j)} \Gamma^k + \Gamma^k \Gamma_{(ij)k} \hspace{3mm}
\nonumber \\
&+& 2 {\Gamma^{mn}}_{(i} \Gamma_{j)mn} + {\Gamma^{mn}}_{i} \Gamma_{mnj} \; ,
\label{eq:RicciNOR1}
\end{eqnarray}
where $\Gamma^i := \gamma^{mn} \Gamma^i_{mn}$.

The crucial difference with the ADM formulation is the fact that the
quantities $\Gamma^i$ that appear in the Ricci tensor above are now
promoted to independent quantities and evolved separately.  To find
the evolution equations for the $\Gamma^i$ we first note that from
their definition we have
\begin{equation}
\Gamma^i = - \partial_m \gamma^{im}
- \frac{1}{2} \: \gamma^{im} \partial_m \ln \gamma \; ,
\end{equation}
with $\gamma$ the determinant of $\gamma_{ij}$.  From this one can
easily show, after some algebra, that:
\begin{eqnarray}
\partial_t \Gamma^i - \lb \Gamma^i &=& \gamma^{lm} \partial_l \partial_m \beta^i
- \frac{2}{\gamma^{1/2}} \: \partial_m \left( \alpha K^{im} \gamma^{1/2} \right)
\nonumber \\
&+& \gamma^{im} \partial_m \left( \alpha K \right) \; , \hspace{5mm}
\label{eq:Gammadot1}
\end{eqnarray}
where the Lie derivative of $\Gamma^i$ that appears in the last
expression should be understood as that of a vector:
\begin{equation}
\lb \Gamma^i := \beta^m \partial_m \Gamma^i
- \Gamma^m \partial_m \beta^i \; .
\end{equation}
In fact, one can now add a multiple of the momentum
constraints~\eqref{eq:momentum} to this equation to obtain a final
evolution equation of the form:
\begin{eqnarray}
\partial_t \Gamma^i - \lb \Gamma^i &=&
\gamma^{mn} \partial_m \partial_n \beta^i
- \left[ 2 K^{im} - \gamma^{im} K \right] \nabla_m \alpha \nonumber \\
&-& \alpha \nabla_m \left[ \left( 2 - \xi \right) K^{im}
- \left( 1 - \xi \right) \gamma^{im} K \right] \nonumber \\
&+& 2 \:\alpha K^{mn} \Gamma^i{}_{mn} - 8 \pi \alpha \xi j^i \; ,
\label{eq:Gammadot2}
\end{eqnarray}
with $\xi$ an arbitrary parameter.  The importance of adding a
multiple of the momentum constraints to the evolution equation for
$\Gamma^i$ comes from the fact that, if one chooses a slicing
condition of the Bona-Masso family
\begin{equation}
\partial_t \alpha - \lb \alpha = - \alpha^2 f(\alpha) K \; ,
\label{eq:BonaMasso}
\end{equation}
with $f(\alpha)$ an arbitrary function of $\alpha$, then the NOR
formulation can be shown to be strongly hyperbolic (and thus
well-posed) if one takes $\xi=2$ and $f>0$, or more generally if one
takes $\xi>0$, $f>0$ and $f \neq 1$ (see {\em e.g.}
reference~\cite{Alcubierre08a}).

Instead of using the Bona-Masso slicing condition, one can assume that
the {\em densitized lapse}\/ defined as \mbox{$\tilde{\alpha} :=
  \alpha \gamma^{-f/2}$}, with $f$ a constant, is an {\em a priori}\/
known function of spacetime, $\tilde{\alpha} = F(t,x^i)$.  The same
results about hyperbolicity then follow.

The standard NOR formulation in fact also adds an arbitrary multiple
of the Hamiltonian constraint of the form $\alpha \eta \gamma_{ij} H$
to the evolution equation of $K_{ij}$, with $\eta$ another free
parameter (for $\eta \neq 0$ one finds a new region of parameter space
where the system is also strongly hyperbolic). However, this point is
of no consequence for the discussion that follows, so we will ignore
it from now on.

The evolution equation for $\Gamma^i$ given above is quite general,
but it has the serious disadvantage that it involves quantities that
are not tensors, such as $\Gamma^i_{mn}$ and $\Gamma^i$ itself. In the
next Section we will address this issue.

%%%%%%%%%%%%%%%%%%%%%%%%%%%%
%%%   NOR - CURVILINEAR  %%%
%%%%%%%%%%%%%%%%%%%%%%%%%%%%

\subsection{Curvilinear coordinates}
\label{sec:NOR-curved}

The NOR formulation just described can in principle be used with any
type of coordinates.  However, when dealing with curvilinear
coordinates, that is coordinates that are non-trivial even in flat
space, one can easily find that the conformal connection functions
$\Gamma^i$ are singular at some points and generally do not behave as
a vector would do.  For example, in spherical coordinates the quantity
$\Gamma^r$ turns out to be singular in flat space, while
$\Gamma^\theta$ is non-zero even if we assume spherical symmetry. This
in itself is not necessarily a major problem, as the equations are
quite general and are consistent in any set of coordinates.  However,
dealing with singular quantities numerically can be troublesome, and
also dealing with non-tensor quantities makes it difficult to compare
evolutions done with the same slicing conditions but different spatial
coordinate systems.  It would then seem like a good idea to replace
the non-covariant quantities $\Gamma^i$ with a true vector.

In order to do this we will start from the tensor $\Delta^i{}_{jk}$
defined in equation~\eqref{eq:Deltadef1} above, and furthermore we will
also assume that the background metric is the flat metric in the same
curvilinear coordinates we are considering.  Notice, in particular,
that in Cartesian coordinates we have $\fG^i_{\;jk}=0$, so that in
that case $\Delta^i{}_{jk}$ and $\Gamma^i{}_{jk}$ are identical.

Just as we did before, we will again define the quantities $\Delta^i$
as in~\eqref{eq:Deltadef2}.  We now want to calculate the evolution
equation for $\Delta^i$.  From the definition above we immediately find
\begin{equation}
\partial_t \Delta^i = \partial_t \Gamma^i
- \fG^i{}_{mn} \partial_t \gamma^{mn} \; ,
\end{equation}
where we have used the fact that the flat background does not
evolve. Using now~\eqref{eq:Gammadot2} and the ADM evolution equation
for $\gamma_{ij}$ one can easily find that
\begin{eqnarray}
\partial_t \Delta^i - \lb \Delta^i &=& \gamma^{mn} \partial_m \partial_n \beta^i
+ \gamma^{mn} \lb \fG^i{}_{mn} \nonumber \\
&-& \left[ 2 K^{im} - \gamma^{im} K \right] \nabla_m \alpha \nonumber \\
&-& \alpha \nabla_m \left[ \left( 2 - \xi \right) K^{im}
- \left( 1 - \xi \right) \gamma^{im} K \right] \nonumber \\
&+& 2 \alpha K^{mn} \Delta^i_{mn} \; ,
\end{eqnarray}
where the term $\lb \fG^i{}_{mn}$ must be calculated as if
$\fG^i{}_{mn}$ where a true tensor. In the previous equation
$\Delta^i$ is clearly a vector, and so is $\partial_t \Delta^i$, but
the right hand side is not manifestly covariant since it involves
partial derivatives of the shift and terms containing $\fG^i{}_{mn}$.
However, this can be easily fixed since one can show that, quite
generally,
\begin{eqnarray}
\gamma^{mn} \fnabla_m \fnabla_n \beta^i &=&
\gamma^{mn} \partial_m \partial_n \beta^i + \gamma^{mn} \lb \fG^i{}_{mn}
\nonumber \\
&+& \beta^l \gamma^{mn} \fR^i{}_{mnl} \; ,
\end{eqnarray}
with $\fR^i{}_{mnl}$ the curvature tensor of the background.  Since in
our case the background is flat by construction, we can use the last
result to rewrite the evolution equation for $\Delta^i$ in the
following way
\begin{eqnarray}
\partial_t \Delta^i - \lb \Delta^i &=&
\gamma^{mn} \fnabla_m \fnabla_n \beta^i \nonumber \\
&-& \left[ 2 K^{im} - \gamma^{im} K \right] \nabla_m \alpha \nonumber \\
&-& \alpha \nabla_m \left[ \left( 2 - \xi \right) K^{im}
- \left( 1 - \xi \right) \gamma^{im} K \right] \nonumber \\
&+& 2 \alpha K^{mn} \Delta^i_{mn} \; .
\label{eq:Deltadot_NOR}
\end{eqnarray}
The last equation is now manifestly covariant.

In summary, in order to use the NOR formulation in curvilinear
coordinates we need to express the \linebreak 3-dimensional Ricci
tensor that appears in the evolution equations for the extrinsic
curvature as (confront this with eq.~\eqref{eq:RicciNOR1}):
\begin{eqnarray}
R_{ab} &=& - \frac{1}{2} \: \gamma^{mn} \fnabla_m \fnabla_n \gamma_{ab}
+ \gamma_{m(a} \fnabla_{b)} \Delta^m + \Delta^m \Delta_{(ab)m} \nonumber \\
&+& 2 \Delta^{mn}{}_{(a} \Delta_{b)mn} + {\Delta^{mn}}_a \Delta_{mnb} \; ,
\label{eq:RicciNOR2}
\end{eqnarray}
with $\Delta^a{}_{bc}$ and $\Delta^a$ defined in~\eqref{eq:Deltadef1}
and~\eqref{eq:Deltadef2}, promote the $\Delta^a$ to independent
quantities, and evolve them through~\eqref{eq:Deltadot_NOR}.

%%%%%%%%%%%%%%%%%%%%%%%%%
%%%   BSSN - GENERAL  %%%
%%%%%%%%%%%%%%%%%%%%%%%%%

\section{The BSSN formulation}
\label{sec:BSSN-general}

\subsection{Standard formulation}
\label{sec:BSSN-standard}

The BSSN formulation is a reformulation of the ADM evolution equations,
based on the work of Shibata and Nakamura~\cite{Shibata95} and
Baumgarte and Shapiro~\cite{Baumgarte:1998te}, that has proven to be
particularly robust in the numerical evolution of a large variety of
spacetimes.  This formulation is based on a conformal decomposition of
the metric of the form
\begin{equation}
\tg_{ij} = e^{- 4 \phi} \gamma_{ij} \; ,
\label{eq:gammatildedef}
\end{equation}
where the conformal factor $\phi$ is chosen in such a way that the
determinant of the conformal metric is unity $\tg=1$,
which implies:
\begin{equation}
\phi = \frac{1}{12} \: \ln \gamma \; .
\label{eq:phidef}
\end{equation}
From the definition above and the ADM evolution equation
for the spatial metric~\eqref{eq:gammadot}, one can easily
find the following evolution equation for $\phi$:
\begin{equation}
\partial_t \phi = - \frac{1}{6} \left( \alpha K
- \partial_m \beta^m \right)
+ \beta^m \partial_m \phi \; .
\label{eq:phidot1}
\end{equation}
The last equation can in fact be rewritten as
\begin{equation}
\partial_t \phi - \lb \phi = - \frac{1}{6} \: \alpha K \; ,
\label{eq:phidot2}
\end{equation}
where the Lie derivative of $\phi$ is given by
\begin{equation}
\lb \phi = \beta^m \partial_m \phi + \frac{1}{6} \: \partial_m \beta^m \; .
\end{equation}
Notice that, strictly speaking, $\phi$ is not a true scalar density
since its definition involves a logarithm, but \mbox{$\psi := e^\phi =
  \gamma^{1/12}$} is a well defined scalar density of weight $1/6$, so
that the Lie derivative of $\phi$ is just \mbox{$\lb \phi = \lb \psi /
  \psi$}, which reduces to the expression above.

The BSSN formulation also separates the extrinsic curvature into its
trace $K$ and its trace-free part
\begin{equation}
A_{ij} = K_{ij} - \frac{1}{3} \: \gamma_{ij} K \; .
\label{eq:defA}
\end{equation}
We further make a conformal rescaling of the traceless extrinsic
curvature of the form
\begin{equation}
\tA_{ij} = e^{-4 \phi} A_{ij}
= e^{-4 \phi} \left( K_{ij} - \frac{1}{3} \: \gamma_{ij} K \right) \; .
\end{equation}

Just as we did in the case of the NOR formulation, the BSSN
formulation also introduces three auxiliary
variables known as the {\em conformal connection functions}\/ and
defined as
\begin{equation}
\tG^i := \tg^{jk} \tG^i_{jk} = - \partial_j \tg^{ij} \; ,
\label{eq:defG}
\end{equation}
where $\tG^i{}_{jk}$ are the Christoffel symbols of the conformal
metric, and where the second equality comes from the fact that the
determinant $\tg$ is equal to 1.

The evolution equation for $\phi$ was already found above, while those
for $\tg_{ij}$, $K$ and $\tA_{ij}$ can be obtained directly from the
standard ADM equations.  The system of evolution equations then takes
the form
\begin{eqnarray}
\partial_t \tg_{ij} - \lb \tg_{ij} &=& - 2 \alpha \tA_{ij} \: ,
\label{eq:dtgdt} \\
\partial_t \phi - \lb \phi &=& - \frac{1}{6} \alpha K \: ,
\label{eq:dphidt} \\
\partial_t \tA_{ij} - \lb \tA_{ij} &=&
e^{-4\phi} \left\{ - \nabla_i \nabla_j \alpha
+ \alpha R_{ij} \right. \nonumber \\
&+& \left. 4 \pi \alpha \left[ \gamma_{ij} \left( S - \rho \right)
- 2 S_{ij} \right] \right\}^{\rm TF} \nonumber \\
&+& \alpha \left( K \tA_{ij} - 2 \tA_{ik} \tA^k{}_j \right) \: ,
\label{eq:dtAdt} \\
\partial_t K - \lb K &=& - \nabla^2 \alpha
+ \alpha \left( \tA_{ij} \tA^{ij} + \frac{1}{3} K^2 \right) \nonumber \\
&+& 4 \pi \alpha \left( \rho + S \right ) \: ,
\label{eq:dtrKdt}
\end{eqnarray}
with $\nabla^2 := \nabla^m \nabla_m$ the spatial Laplacian operator
associated with the full physical metric, and where TF denotes the
trace-free part of the expression inside the brackets.  Notice also
that indices of conformal quantities are assumed to be raised and
lowered with the conformal metric.  Here it is important to mention
that the Hamiltonian constraint has already been used in the evolution
equation for $K$ in order to eliminate the Ricci scalar.

In the evolution equation for $\tA_{ij}$ above one needs to calculate
the Ricci tensor associated with the physical metric, which can be
separated into two contributions in the following
way:
\begin{equation}
R_{ij} = \tR_{ij} + R^{\phi}_{ij} \; ,
\end{equation}
where $\tR_{ij}$ is the Ricci tensor associated with the
conformal metric $\tg_{ij}$, which we write in terms of the
$\tG^i$ as
\begin{eqnarray}
\tilde{R}_{ij} &=& - \frac{1}{2} \tg^{mn} \partial_m \partial_n
\tg_{ij} + \tg_{k(i} \partial_{j)} \tG^k + \tG^k \tG_{(ij)k} \nonumber \\
&+& 2 \tG^{mn}{}_{(i} \tG_{j)mn} + \tG^{mn}{}_{i} \tG_{mnj} \; ,
\label{eq:Ricciconf}
\end{eqnarray}
(this is just the standard expression~\eqref{eq:Ricci} for the
conformal metric), and where $R^{\phi}_{ij}$ denotes additional terms
that depend on derivatives of $\phi$:
\begin{eqnarray}
R^{\phi}_{ij} &=& - 2 \tnabla_i \tnabla_j \phi
- 2 \tg_{ij} \tnabla^k \tnabla_k \phi \nonumber \\
&+& 4 \tnabla_i \phi \: \tnabla_j \phi
- 4 \tg_{ij} \tnabla^k \phi \: \tnabla_k \phi \; ,
\label{eq:Ricciphi}
\end{eqnarray}
with $\tnabla_i$ the covariant derivative associated with the
conformal metric.

Notice also that the evolution equations for $\tA_{ij}$ and $K$
involve covariant derivatives of the lapse function with respect to
the physical metric $\gamma_{ij}$ ({\em i.e.} covariant derivatives
with no tilde).  One must also be careful with the fact that in the
evolution equations above we need to calculate Lie derivatives with
respect to the shift vector $\beta^i$ of tensor densities.  In
particular, $\tg_{ij}$ and $\tA_{ij}$ are tensor densities of weight
$-2/3$.

We are still missing an evolution equation for the $\tG^i$.  This
equation can be obtained directly from the definition,
equation~\eqref{eq:defG}.  One finds:
\begin{eqnarray}
\partial_t \tG^i - \lb \tG^i &=& \tg^{jk} \partial_j \partial_k \beta^i
+ \frac{1}{3} \: \tg^{ij} \partial_j \partial_k \beta^k \nonumber \\
&-& 2 \left( \alpha \partial_j \tA^{ij}
+ \tA^{ij} \partial_j \alpha \right) \; .
\label{eq:dtGdt0}
\end{eqnarray}
In the above equation the Lie derivative of $\tG^i$ should be
calculated as if $\tG^i$ where a vector density of weight $2/3$:
\begin{equation}
\lb \tG^i = \beta^j \partial_j \tG^i - \tG^j \partial_j \beta^i
+ \frac{2}{3} \: \tG^i \partial_j \beta^j \; .
\end{equation}

Again, just as we did in the case of NOR, we will modify the
evolution equation for $\tG^i$ given above by adding to it
a multiple of the momentum constraints. In order to do this,
let us first rewrite the Hamiltonian and momentum constraints
in terms of the conformally rescaled quantities.  One finds:
\begin{eqnarray}
H &:=& \frac{1}{2} \left( R + \frac{2}{3} K^2 - \tA_{ij} \tA^{ij} \right)
- 8 \pi \rho = 0 \; , \qquad \label{eq:hamiltonianBSSN} \\
M^i &:=& \nabla_j A^{ij} - \frac{2}{3} \gamma^{ij} \partial_j K
- 8 \pi j^i \nonumber \\
&=& e^{- 4 \phi} \left( \tilde{\nabla}_j \tA^{ij} - \frac{2}{3}
\tg^{ij} \partial_j K + 6 \tA^{i} \partial_j \phi \right) \nonumber \\
&-& 8 \pi j^i = 0 \; . \label{eq:momentumBSSN}
\end{eqnarray}
where $\tilde{\nabla}_i$ now denotes covariant derivative with respect
to the conformal metric.  Notice also that the fact that the covariant
metric has unit determinant implies that the term $\tilde{\nabla}_j
\tA^{ij}$ can be written as:
\begin{equation}
\tilde{\nabla}_j \tA^{ij} = \partial_j \tA^{ij} + \tG^i{}_{jk} \tA^{jk} \; .
\end{equation}

The Hamiltonian constraint above was in fact already used in order to
eliminate the Ricci scalar from the evolution equation for the trace
of the extrinsic curvature $K$ above (equation~\eqref{eq:dtrKdt}).

Adding now a multiple of the momentum constraint to the evolution
equation for $\tG^i$, equation~\eqref{eq:dtGdt0} above, we find:
\begin{eqnarray}
\partial_t \tG^i - \lb \tG^i &=& \tg^{jk} \partial_j \partial_k \beta^i
+ \frac{1}{3} \: \tg^{ij} \partial_j \partial_k \beta^k
- 2 \tA^{ij} \partial_j \alpha \nonumber \\
&-& \alpha \left( 2 - \xi \right) \partial_j \tA^{ij}
+ \alpha \xi \left( \tG^i_{jk} \tA^{jk} \right. \nonumber \\
&+& \left. 6 \tA^{ij} \partial_j \phi - \frac{2}{3} \tg^{ij} \partial_j K
- 8 \pi \tilde{j}^i \right) .
\label{eq:dtGdt1}
\end{eqnarray}
with $\xi$ an arbitrary parameter, and where $\tilde{j}^i := e^{4
  \phi} j^i$.  The standard BSSN formulation usually takes $\xi=2$,
which seems to be an optimal choice.

Just as in the case of the NOR formulation, The BSSN formulation just
described can be shown to be strongly hyperbolic for
$\xi>1/2$~\cite{Alcubierre08a}.  Standard BSSN with $\xi=2$ has turned
out to be particularly robust in practice, and leads to stable and
well behaved numerical simulations.  In conjunction with the
Bona-Masso slicing condition~\eqref{eq:BonaMasso}, and the so-called
``Gamma driver'' shift condition~\cite{Alcubierre02a} (see
Section~\ref{sec:gammadriver} below), it has allowed for the accurate
simulation of the inspiral collision of black holes with different
masses and
spins~\cite{Campanelli:2005dd,Campanelli:2006gf,Baker:2005vv,Baker:2006yw}.
Today, most 3-dimensional production numerical relativity codes use
the BSSN formulation in one way or another, the notable exception
being codes that use the ``generalized harmonic
formulation'' (see {\em e.g.} ~\cite{Pretorius:2004jg}).

%%%%%%%%%%%%%%%%%%%%%%%%%%%%%
%%%   BSSN - CURVILINEAR  %%%
%%%%%%%%%%%%%%%%%%%%%%%%%%%%%

\subsection{Curvilinear coordinates}
\label{sec:BSSN-curved}

When adapting the standard BSSN formulation to curvilinear coordinates
we are faced with two problems. The first one is essentially the same
problem that we had with the NOR formulation, namely that the
quantities $\tG^i$ are not vectors (or more specifically vector
densities, but we will come back to that point below).  The second
problem is the fact that in curvilinear coordinates the determinant of
the flat metric is generally different from unity, so that asking for
$\tg=1$ is not a good idea.  Consider, for example, flat space in
spherical coordinates $(r,\theta,\varphi)$ for which the spatial metric
is:
\begin{equation}
ds^2 = dr^2 + r^2 d \Omega^2 \; ,
\end{equation}
with $d \Omega^2 = d \theta^2 + \sin^2 \theta d \varphi^2$ the
standard solid angle element. We then find that \mbox{$\gamma = r^4
  \sin^2 \theta$}.

In curvilinear coordinates it is in fact much better to ask for the
determinant of the conformal metric to reduce to its value in flat
space (see {\em e.g.}~\cite{Brown:2009dd}).  In order to avoid
confusion between the conformal metric of the standard BSSN
formulation and the one we will use here, from now on we will denote
conformal quantities with a hat instead of a tilde.  Also, for the
conformal factor we will use $\chi$ instead of $\phi$, so that we will
in fact have
\begin{equation}
\hg_{ij} = e^{- 4 \chi} \gamma_{ij} \; ,
\label{eq:gammatildecurvedef}
\end{equation}
and we will ask for $\hg(t=0) = \fg$, with $\fg$ the determinant of
the flat metric background in the same curvilinear coordinates.  This
change introduces two new features into the BSSN formulation.  In the
first place, in general we will find that $\hg$ will not be constant
in space so that we can no longer ignore its spatial derivatives.  But
more importantly, it is now not immediately clear how $\hg$ should
evolve in time.

Following Brown~\cite{Brown:2009dd}, one can suggest at least two
``natural'' choices for the evolution of $\hg$:

\begin{enumerate}

\item $\partial_t \hg = 0$.  This is called a ``Lagrangian''
  condition since the determinant of the conformal metric is constant
  along time lines.

\item $\partial_t \hg - \lb \hg = 0$.  This is instead an ``Eulerian''
  condition, since the determinant of the conformal metric is now
  constant along the normal lines ({\em i.e.} it remains constant in
  time as seen by the Eulerian observers), so that it can in fact
  evolve along the time lines.

\end{enumerate}

Standard BSSN then corresponds to the Lagrangian case in
Cartesian coordinates.  Using now the fact that:
\begin{equation}
\lb \hg = \beta^m \partial_m \hg + 2 \hg \partial_m \beta^m
= 2 \hg \hnabla_m \beta^m \; ,
\label{eq:Liegamma}
\end{equation}
we can write in general for the evolution of $\hg$:
\begin{equation}
\partial_t \hg = s \left( 2 \hg \hnabla_m \beta^m \right) \: ,
\label{eq:detgdotcurve}
\end{equation}
with:
\begin{equation}
s = \left\{
\begin{array}{ll}
0 & \quad \text{Lagrangian} \;, \\
1 & \quad \text{Eulerian} \;.
\end{array}
\right.
\end{equation}

On the other hand, since now $\hg \neq 1$, we find for the
conformal factor $\chi$:
\begin{equation}
\chi = \frac{1}{12} \: \ln \left( \gamma / \hg \right) \; .
\label{eq:chidefcurve}
\end{equation}
We can now use this to find the evolution equation for $\chi$:
\begin{eqnarray}
\partial_t \chi &=& \frac{1}{12} \left( \frac{\partial_t \gamma}{\gamma}
- \frac{\partial_t \hg}{\hg} \right) \nonumber \\
&=& \frac{1}{12} \left( - 2 \alpha K + \frac{\lb \gamma}{\gamma} 
- s \frac{\lb \hg}{\hg} \right) \; ,
\end{eqnarray}
which implies:
\begin{eqnarray}
\partial_t \chi - \lb \chi &=& - \frac{1}{6} \alpha K
+ \frac{1}{6} \left( 1 - s \right) \hnabla_m \beta^m \nonumber \\
&=& - \frac{1}{6} \alpha K + \frac{1}{6} \sigma \hnabla_m \beta^m \; ,
\label{eq:chidotcurve}
\end{eqnarray}
where we have used equation~\eqref{eq:Liegamma} above, and where
\mbox{$\lb \chi := \lb \gamma / \gamma - \lb \hg / \hg$}.  In the
above equation we have also introduced the shorthand $\sigma=(1-s)$,
so that $\sigma=1$ now corresponds to a Lagrangian evolution and
$\sigma=0$ to an Eulerian evolution.

There is an important point regarding the tensorial character of $\chi$
that should be mentioned here.  Notice that because of the new
definition of $\chi$, equation~\eqref{eq:chidefcurve}, we now have:
\begin{eqnarray}
\lb \chi &=& \frac{1}{12} \left( \frac{\lb \gamma}{\gamma}
- \frac{\lb \hg}{\hg} \right)
\nonumber \\
&=& \frac{1}{6} \left( \nabla_m \beta^m - \hnabla_m \beta^m \right)
\nonumber \\
&=& \frac{1}{12} \: \beta^m \partial_m \ln \left( \gamma / \hg \right) \; ,
\end{eqnarray}
so that finally
\begin{equation}
\lb \chi = \beta^m \partial_m \chi \; .
\end{equation}
In other words, the Lie derivative of $\chi$ is now that of a scalar
function with {\em no density weight}. We will see below that this
will be the case for all dynamical quantities.  This is another
important difference between the standard BSSN formulation and the
generalization we are introducing here, and it can be traced back to
the fact that the definition~\eqref{eq:chidefcurve} of the conformal
factor $\chi$ now involves the ratio of two volume elements, so that
$\chi$ is a true scalar.

The next step is to find the evolution equation for $\hg_{ij}$.
Starting from the definition~\eqref{eq:gammatildecurvedef} above, and
using the evolution equation for $\chi$~\eqref{eq:chidotcurve},
together with the ADM evolution equation for $\gamma_{ij}$ given
by~\eqref{eq:gammadot}, we now find
\begin{equation}
\partial_t \hg_{ij} - \lb \hg_{ij} = - 2 \alpha \hA_{ij}
- \frac{2}{3} \sigma \: \hg_{ij} \hnabla_m \beta^m \: ,
\label{eq:BSSNgammadotcurve}
\end{equation}
where $\hA_{ij}$ is now defined as:
\begin{equation}
\hA_{ij} := e^{- 4 \chi} \left( K_{ij} - \frac{1}{3} \: \gamma_{ij} K 
\right) \: .
\end{equation} 
Again, with the definition of $\chi$ above, both $\hg_{ij}$ and
$\hA_{ij}$ are true tensors and not tensor densities, and their
Lie derivatives should be calculated accordingly.

Similarly, the evolution equations for $\hA_{ij}$ and $K$ become:
\begin{eqnarray}
\partial_t \hA_{ij} - \lb \hA_{ij} &=& e^{-4\chi}
\left\{ - \nabla_i \nabla_j \alpha
+ \alpha R_{ij} \right. \nonumber \\
&+& \left. 4 \pi \alpha \left[ \gamma_{ij} \left( S - \rho \right)
- 2 S_{ij} \right] \right\}^{\rm TF} \nonumber \\
&+& \alpha \left( K \hA_{ij} - 2 \hA_{ik} \hA^k{}_j \right) \nonumber \\
&-& \frac{2}{3} \sigma \: \hA_{ij} \hnabla_m \beta^m \: ,
\label{eq:dtAdt2} \\
\partial_t K - \lb K &=&  - \nabla^2 \alpha
+ \alpha \left( \hA_{ij} \hA^{ij} + \frac{1}{3} K^2 \right) \nonumber \\
&+& 4 \pi \alpha \left( \rho + S \right ) \: ,
\label{eq:dtrKdt2}
\end{eqnarray}
Again, the Lie derivatives on the left-hand side are now those of
proper tensors with no density weight.  Notice that there is no term
with $\sigma \hnabla_m \beta^m$ in the evolution equation for $K$
since it is a scalar.

Just as before, the Ricci tensor that appears in the evolution
equation for $\hA_{ij}$ is now separated into two contributions in the
following way
\begin{equation}
R_{ij} = \hR_{ij} + R^{\chi}_{ij} \; ,
\end{equation}
where $\hR_{ij}$ is the Ricci tensor associated with the conformal
metric $\hg_{ij}$, and where $R^{\chi}_{ij}$ denotes the terms that
depend on derivatives of $\chi$:
\begin{eqnarray}
R^{\chi}_{ij} &=& - 2 \hnabla_i \hnabla_j \chi
- 2 \hg_{ij} \hnabla^k \hnabla_k \chi \nonumber \\
&+& 4 \hnabla_i \chi \: \hnabla_j \chi
- 4 \hg_{ij} \hnabla^k \chi \: \hnabla_k \chi \; ,
\label{eq:Riccichi}
\end{eqnarray}
with $\hnabla_i$ the covariant derivative associated with the
conformal metric $\hg_{ij}$.

Following what we did in the case of the NOR formulation, we now want
to write a fully covariant expression for the conformal Ricci tensor
$\hR_{ij}$.  In order to do so we will again introduce the quantities
\begin{eqnarray}
\hD^a{}_{bc} &:=& \hG^a{}_{bc} - \fG^a{}_{bc} \; , \\
\hD^i &:=& \hg^{mn} \hD^i{}_{mn} = \hG^i - \hg^{mn} \fG^i{}_{mn} \; .
\label{eq:BSSN-Deltadef}
\end{eqnarray}
With these definitions the conformal Ricci tensor can be written as
\begin{eqnarray}
\hR_{ab} &=& - \frac{1}{2} \: \hg^{mn} \fnabla_m \fnabla_n \hg_{ab}
+ \hg_{m(a} \fnabla_{b)} \hD^m + \hD^m \hD_{(ab)m} \nonumber \\
&+& 2 \hD^{mn}{}_{(a} \hD_{b)mn} + {\hD^{mn}}_a \hD_{mnb} \; ,
\label{eq:RicciBSSN2}
\end{eqnarray}

The next step is to promote the $\hD^i$ to independent variables and
find an evolution equation for them.  In order to do this we must
first find the evolution equation for the $\hG^i$. Notice that, since
now we have $\hg \neq 1$, the $\hG^i$ now take the form
\begin{equation}
\hG^i = \hg^{mn} \hG^i{}_{mn}
= - \partial_m \hg^{im} - \frac{1}{2} \: \hg^{im} \partial_m \ln \hg \; .
\end{equation}
Using equations~\eqref{eq:detgdotcurve}
and~\eqref{eq:BSSNgammadotcurve} we now find, after some algebra:
\begin{eqnarray}
\partial_t \hG^i &=& \lb \hG^i + \hg^{mn} \partial_m \partial_n \beta^i
- \frac{2}{\hg^{1/2}} \: \partial_m \left( \alpha \hA^{im} \hg^{1/2} \right)
\nonumber \\
&+& \frac{\sigma}{3} \left[ \hg^{im} \partial_m \left(
\hnabla_n \beta^n \right) 
+ 2 \hG^i \: \hnabla_n \beta^n \right] , \qquad
\end{eqnarray}
where $\lb \hG^i$ is calculated as the Lie derivative of a vector:
\begin{equation}
\lb\hG^i = \beta^m \partial_m \hG^i - \hG^m \partial_m \beta^i \; .
\end{equation}

The evolution equation for $\hD^i$ can now be obtained from the last
equation using the fact that
\begin{equation}
\partial_t \hD^i = \partial_t \hG^i - \fG^i{}_{mn} \partial_t \hg^{mn} \; ,
\end{equation}
where, just as we did in the case of the NOR formulation, we have
assumed that the flat background does not evolve.  One finds
\begin{eqnarray}
\partial_t \hD^i &=& \lb \hD^i + \hg^{mn} \partial_m \partial_n \beta^i
+ \hg^{mn} \lb \fG^i{}_{mn} \nonumber \\
&-& 2 \hnabla_m \left( \alpha \hA^{im} \right)
+ 2 \alpha \hA^{mn} \hD^i{}_{mn} \nonumber \\
&+& \frac{\sigma}{3} \left[ \hnabla^i \left( \hnabla_n \beta^n \right)
+ 2 \hD^i \: \hnabla_n \beta^n \right] , \qquad
\end{eqnarray}
where again the Lie derivative of $\hD^i$ is that of a true vector
with no density weight, and the term $\lb \fG^i{}_{mn}$ should be
calculated as the Lie derivative of a tensor.

Just as it happened in the case of the NOR formulation, the right hand
side of the last equations contains terms that involve partial
derivatives of the shift, and also terms containing $\fG^i{}_{mn}$, so
the expression is not explicitly covariant.  We can again fix this by
using the fact that on a flat background the following relation holds:
\begin{equation}
\hg^{mn} \fnabla_m \fnabla_n \beta^i =
\hg^{mn} \partial_m \partial_n \beta^i + \hg^{mn} \lb \fG^i{}_{mn} \; ,
\end{equation}
so that the evolution equation for $\hD^i$ takes the final form
\begin{eqnarray}
\partial_t \hD^i &-& \lb \hD^i \,\,=\,\, \hg^{mn} \fnabla_m \fnabla_n \beta^i
\nonumber \\
&-& 2 \hnabla_m \left( \alpha \hA^{im} \right)
+ 2 \alpha \hA^{mn} \hD^i{}_{mn} \nonumber \\
&+& \frac{\sigma}{3} \left[ \hnabla^i \left( \hnabla_n \beta^n \right)
+ 2 \hD^i \: \hnabla_n \beta^n \right] , \qquad
\end{eqnarray}
which is now manifestly covariant.

The last step is to add a multiple of the momentum constraints to
the evolution equation for $\hD^i$ above.  Doing that we finally
find:
\begin{eqnarray}
\partial_t \hD^i &-& \lb \hD^i \,\,=\,\, \hg^{mn} \fnabla_m \fnabla_n \beta^i
- 2 \hA^{im} \partial_m \alpha \nonumber \\
&-& \alpha \left( 2 - \xi \right) \hnabla_m \tA^{im}
+ 2 \alpha \hA^{mn} \hD^i{}_{mn} \nonumber \\
&+& \alpha \xi \left( 6 \tA^{ij} \partial_j \phi
- \frac{2}{3} \tg^{ij} \partial_j K - 8 \pi \tilde{j}^i \right) \nonumber \\
&+& \frac{\sigma}{3} \left[ \hnabla^i \left( \hnabla_n \beta^n \right)
+ 2 \hD^i \: \hnabla_n \beta^n \right] , \qquad
\label{eq:Deltadot}
\end{eqnarray}
where again $\xi$ is an arbitrary constant that must be such that
$\xi>1/2$ for the final system to be strongly hyperbolic.

\vspace{5mm}

We can now ask which would be the preferred choice for $\sigma$ in the
above evolution equations, that is, should we take a Lagrangian or an
Eulerian approach?  Looking at the evolution equation for $\chi$,
equation~\eqref{eq:chidotcurve}, one might first think that the
simplest choice would be to take \mbox{$\sigma=0$} ($s=1$), that is an
Eulerian approach, since in that case the evolution equation
simplifies.  However, from the discussion above about the scalar
character of $\chi$ we see that if we choose $\sigma=0$, the evolution
equation for $\chi$ {\em does not reduce}\/ to the standard BSSN
evolution equation for $\phi$ given by equation~\eqref{eq:phidot2} in
the case of Cartesian coordinates (for which $\hg=1$).  This statement
might seem somewhat puzzling since for $\sigma=0$
equations~\eqref{eq:phidot2} and~\eqref{eq:chidotcurve} look
identical.  However, one must remember that $\chi$ is a true scalar,
while $\phi$ is a scalar density, so that their Lie derivatives are
different.  The same is true for the evolution equations of
$\hg_{ij}$, $\hA_{ij}$ and $\hD^i$.

It is in fact not difficult to convince oneself that if we want to
recover the standard BSSN evolution equations in the case of Cartesian
coordinates we must choose $\sigma=1$, {\em i.e.}\/ the Lagrangian
approach, and simply remember that all dynamical quantities are now
true tensors with {\em no}\/ density weight. The terms corresponding
to the Lie derivatives of tensor densities in standard BSSN now appear
explicitly on the right-hand side of the evolution equations through
the terms proportional to \mbox{$\hnabla_m \beta^m$}.

%%%%%%%%%%%%%%%%%%%%%%%%%%%%%%
%%%   GAMMA DRIVER SHIFT   %%%
%%%%%%%%%%%%%%%%%%%%%%%%%%%%%%

\subsection{The Gamma driver shift condition}
\label{sec:gammadriver}

The equations presented in the previous Section are completely
general, and can be used with any gauge condition, both for the lapse
and the shift.  One we mentioned the issue of hyperbolicity we
specified a particular slicing condition (the Bona-Masso condition),
and the use of a shift known {\em a priori}, but this was just in
order to make the discussion concrete.

One particularly important shift condition that has turned out to be
extremely robust in practice, and in the last few years has allowed
for the stable and accurate simulation of inspiraling black holes is
the so-called ``Gamma driver'' shift condition~\cite{Alcubierre02a}.
This shift condition is particularly well-adapted to the BSSN
formulation, and comes in two versions (with several variations).  The
first possibility is the ``parabolic'' Gamma driver which takes the
form:
\begin{equation}
\partial_t \beta^i = c_1 \partial_t \hG^i \; ,
\end{equation}
with $c_1$ some positive constant.  The reason for the name
``parabolic'' is that, since the time derivative of $\hG^i$ that
appears in the right-hand side involves second derivatives of the
shift, the above equation results in a generalized heat-like equation
for the shift components.  In numerical simulations, the above
condition has the same problem as any other parabolic equation, namely
that for numerical stability the time step must be proportional to the
square of the spatial grid spacing, resulting in a prohibitive use of
computational resources in the case of high resolution simulations.

The second version of the Gamma driver is the so-called ``hyperbolic''
Gamma driver which can be written in two alternative forms.  The first
form is simply
\begin{equation}
\partial_t \beta^i = c_2 \hG^i \; ,
\end{equation}
while the second is
\begin{equation}
\partial^2_t \beta^i = c_2 \partial_t \hG^i \; .
\end{equation}
In both cases we end up with a generalized wave equation for the shift,
which justifies the name ``hyperbolic''.  The second version is
usually preferred since it allows one to add a damping term that has been
found to be very important in numerical simulations:
\begin{equation}
\partial^2_t \beta^i = c_2 \partial_t \hG^i - \eta \partial_t \beta^i \; .
\end{equation}
For reasons that we will not go into here, typical values of the
parameters are $c_2 = 3/4$ and $\eta = 2/M_{\rm ADM}$ (notice that
$\eta$ has dimensions of inverse distance, so it is usually
scaled with the total ADM mass of the spacetime).

The main problem with the above shift conditions from the point of
view of our present discussion is that, while they work well in
Cartesian coordinates, they are seriously flawed in curvilinear
coordinates since on the left-hand side we have a proper vector, while
on the right-hand side we have contracted Christoffel symbols. But we
see that this problem can now be easily solved by choosing conditions
of the form
\begin{eqnarray}
\partial_t \beta^i = c_1 \partial_t \hD^i && \quad {\rm parabolic} \; , \\
\partial^2_t \beta^i = c_2 \partial_t \hD^i - \eta \partial_t \beta^i
&& \quad {\rm hyperbolic} \; .
\end{eqnarray}
When working in curvilinear coordinates, one must then use these
modified Gamma driver conditions (or rather ``Delta driver''
conditions) in order to keep everything consistent.

%%%%%%%%%%%%%%%%%%%%%%%%%%%%%%
%%%   SPHERICAL SYMMETRY   %%%
%%%%%%%%%%%%%%%%%%%%%%%%%%%%%%

\section{The case of spherical symmetry}
\label{sec:spherical}

Having found the general form of the NOR and BSSN equations in
curvilinear coordinates, we will now consider the special case of
spherical symmetry.  Here we will concentrate on the case of the BSSN
formulation, for the NOR formulation the analysis is entirely
analogous (and in fact somewhat simpler).

%%%%%%%%%%%%%%%%%%%%%%%%%%%%%%%%%%%%%%
%%%   BSSN IN SPHERICAL SYMMETRY   %%%
%%%%%%%%%%%%%%%%%%%%%%%%%%%%%%%%%%%%%%

\subsection{BSSN in spherical symmetry}

\subsubsection{Main equations}
\label{sec:BSSNsphere}

We start by writing the general form of the spatial metric in
spherical symmetry as
\begin{equation}
dl^2 = e^{4 \chi} \left( a(r,t) dr^2 + r^2 b(r,t) d \Omega^2 \right) \; ,
\label{eq:spheremetric}
\end{equation}
with $a(r,t)$ and $b(r,t)$ positive metric functions, $d \Omega^2$ the
solid angle element $d \Omega^2 = d \theta^2 + \sin^2 \theta d
\varphi^2$, and where $\chi$ is the BSSN conformal factor introduced
in Section~\ref{sec:BSSN-curved} above.  Notice that with this
notation the components of the conformal metric are $\hg_{rr}=a$,
\mbox{$\hg_{\theta \theta}=r^2 b$}, and \mbox{$\hg_{\varphi
    \varphi}=(r \sin \theta)^2 \: b$}.

The determinants of the physical and conformal metric take the form:
\begin{eqnarray}
\gamma &=& a b^2 \left( r^4 e^{12 \chi} \sin^2 \theta \right) \; , \\
\hg &=& a b^2 \left( r^4 \sin^2 \theta \right)  \; .
\end{eqnarray}
The determinant of the flat metric in spherical coordinates can be 
easily found by setting $a=b=1$ in the expression for $\hg$ above:
\begin{equation}
\fg = r^4 \sin^2 \theta \; .
\end{equation}
The condition that $\hg(t=0) = \fg$ now implies that initially we must
ask for $a b^2 = 1$.

Notice in particular that for a Lagrangian evolution ($\sigma=1$) the
metric components $a$ and $b$ are in fact not independent of each
other.  This is because in that case the determinant of the conformal
metric $\hg$ remains constant in time, so that the relation $a b^2=1$
will always hold.  In the Eulerian case ($\sigma=0$) the quantity $a
b^2$ does evolve, but its evolution is entirely controlled by the
shift, and is independent of both the lapse and the extrinsic
curvature.

Let us now consider the shift vector.  Since we are in spherical
symmetry, the shift (as well as any other vector) will only have a
radial component: \mbox{$\beta^i=(\beta^r,0,0)$}.  Since the different
evolution equations in BSSN involve the conformal divergence of the
shift it is convenient at this point to calculate it. One finds, after
some algebra:
\begin{eqnarray}
\hnabla_m \beta^m &=& \partial_r \beta^r
+ \beta^r \left( \frac{\partial_r a}{2a} + \frac{\partial_r b}{b}
+ \frac{2}{r} \right) \nonumber \\
&=& \partial_r \beta^r + \beta^r \left( \frac{\partial–r (a b^2)}{2 a b^2}
+ \frac{2}{r} \right)\; .
\label{eq:sphere-divbeta}
\end{eqnarray}

Consider next the auxiliary vector $\hD^i$.  Since this is a true vector
(this is the whole idea), it will again only have a radial component:
\mbox{$\hD^i=(\hD^r,0,0)$}. One can show that this in indeed the case
from the definition~\eqref{eq:BSSN-Deltadef}. For the radial component we find
\begin{equation}
\hD^r = \frac{1}{a} \left[ \frac{\partial_r a}{2a}
- \frac{\partial_r b}{b} - \frac{2}{r} \left( 1
- \frac{a}{b} \right) \right] \; .
\label{eq:sphere-Delta}
\end{equation}
One must remember, however, that in what follows $\hD^r$ will be
promoted to an independent variable and the equation above will be
considered a constraint.

Let us now find the specific form of the evolution equation for the
conformal factor $\chi$ and the components of the conformal metric $a$
and $b$.  From equation~\eqref{eq:chidotcurve} we find
\begin{equation}
\partial_t \chi = \beta^r \partial_r \chi + \sigma \hnabla_m \beta^m
- \frac{1}{6} \alpha K \; ,
\label{eq:sphere-chidot}
\end{equation}
where the divergence of the shift is given
by~\eqref{eq:sphere-divbeta} above.  For the conformal metric
components we find
\begin{eqnarray}
\partial_t a &=& \beta^r \partial_r a + 2 a \partial_r \beta^r
- \frac{2}{3} \: \sigma a \: \hnabla_m \beta^m - 2 \alpha a A_a , \qquad
\label{eq:sphere-adot} \\
\partial_t b &=& \beta^r \partial_r b + 2 b \: \frac{\beta^r}{r}
- \frac{2}{3} \: \sigma b \: \hnabla_m \beta^m - 2 \alpha b A_b \; .
\label{eq:sphere-bdot}
\end{eqnarray}
where we have introduced the quantities
\begin{equation}
A_a := \hA^r_r \; , \qquad A_b := \hA^\theta_\theta \; .
\end{equation}

The reason for using the mixed components of the traceless extrinsic
curvature instead of the fully covariant ones is that in spherical
symmetry such a choice simplifies considerably the evolution
equations. In particular, the fact that the tensor $\hA_{ij}$ must be
traceless implies that
\begin{equation}
A_a + 2 A_b = 0 \; .
\end{equation}
Notice also that in fact one has
\begin{equation}
\hA^r_r = \hg^{rr} \hA_{rr} = \gamma^{rr} A_{rr} = A^r_r \; ,
\end{equation}
and similarly for the angular component, so when we use mixed
components of second-rank tensors the ``conformal'' and ``physical''
versions are identical.

Consider next the evolution equation for the trace of the extrinsic
curvature $K$.  We find:
\begin{eqnarray}
\partial_t K &=& \beta^r \partial_r K - \nabla^2 \alpha
+ \alpha \left( A_a^2 + 2 A_b^2 + \frac{1}{3} \: K^2\right) \nonumber \\
&+& 4 \pi \alpha \left( \rho + S_a + 2 S_b \right) , \quad
\label{eq:sphere-Kdot}
\end{eqnarray}
with $S_a$ and $S_b$ the mixed components of the stress tensor
\begin{equation}
S_a := S^r_r \; , \qquad S_b := S^\theta_\theta \; ,
\end{equation}
and where the physical Laplacian of the lapse is given by
\begin{eqnarray}
\nabla^2 \alpha &=&  \frac{1}{a e^{4 \chi}} \left[ \partial_r^2 \alpha \right.
\nonumber \\
&-& \left. \partial_r \alpha \left( \frac{\partial_r a}{2a}
- \frac{\partial_r b}{b}
- 2 \partial_r \chi - \frac{2}{r} \right) \right] . \qquad
\label{eq:sphere-lapalpha}
\end{eqnarray}

The evolution equation for the traceless part of the extrinsic
curvature is somewhat more complicated.  Remember first that we only
need an evolution equation for $A_a$, since the traceless condition
implies $A_b = - A_a/2$.  Rewriting equation~\eqref{eq:dtAdt2} for
the case of spherical symmetry we find
\begin{eqnarray}
\partial_t A_a &=& \beta^r \partial_r A_a - \left( \nabla^r \nabla_r \alpha
- \frac{1}{3} \nabla^2 \alpha \right)
+ \alpha \left( R^r_r - \frac{1}{3} R \right) \nonumber \\
&+& \alpha K A_a - 16 \pi \alpha \left( S_a - S_b \right) \; ,
\label{eq:sphere-Adot}
\end{eqnarray}
where $\nabla^2 \alpha$ was given above and
\begin{equation}
\nabla^r \nabla_r \alpha = \frac{1}{a e^{4 \chi}} \left[ \partial_r^2 \alpha 
- \partial_r \alpha \left( \frac{\partial_r a}{2a}
+ 2 \partial_r \chi \right) \right] \; ,
\end{equation}
and where the mixed radial component of the Ricci tensor $R^r_r$ and
its trace $R$ are given by
\begin{eqnarray}
R^r_r &=& - \frac{1}{a e^{4 \chi}} \left[ \frac{\partial^2_r a}{2a} 
- a \partial_r \hD^r  - \frac{3}{4} \left( \frac{\partial_r a}{a} \right)^2
\right. \nonumber \\
&+& \frac{1}{2} \left( \frac{\partial_r b}{b} \right)^2
- \frac{1}{2} \hD^r \partial_r a + \frac{\partial_r a}{rb} \nonumber \\
&+& \frac{2}{r^2} \left( 1 - \frac{a}{b} \right) \left( 1
+ \frac{r \partial_r b}{b} \right)
\nonumber \\
&+& 4 \left. \partial^2_r \chi - 2 \partial_r \chi \left( \frac{\partial_r a}{a}
- \frac{\partial_r b}{b} - \frac{2}{r} \right) \right] ,
\label{eq:sphere-Rrr} \\
R &=& - \frac{1}{a e^{4 \chi}} \left[ \frac{\partial^2_r a}{2a}
+ \frac{\partial^2_r b}{b} - a \partial_r \hD^r
- \left( \frac{\partial_r a}{a} \right)^2 \right. \nonumber \\
&+& \frac{1}{2} \left( \frac{\partial_r b}{b} \right)^2
+ \frac{2}{rb} \left( 3 - \frac{a}{b} \right) \partial_r b \nonumber \\
&+& \frac{4}{r^2} \left( 1 - \frac{a}{b} \right)
+ 8 \left( \partial^2_r \chi + ( \partial_r \chi )^2 \right) \nonumber \\
&-& \left. 8 \partial_r \chi \left( \frac{\partial_r a}{2a}
- \frac{\partial_r b}{b} - \frac{2}{r} \right) \right] . \hspace{10mm}
\label{eq:sphere-RSCAL}
\end{eqnarray}

It is interesting to notice that in equation~\eqref{eq:sphere-Adot}
there is no contribution from the divergence of the shift (and hence
$\sigma$ plays no role). The reason for this is that even though such
terms do appear in the evolution equation for $A_{rr}$, once we raise
the index to find the evolution equation for $A_a = A^r_r$ they
cancel out.  In fact, the shift contribution reduces to a pure
advection term $\beta^r \partial_r A_a$. This is one of the reasons
why working with the mixed components is useful.

Finally, we need an evolution equation for $\hD^r$.
Writing~\eqref{eq:Deltadot} for the case of spherical symmetry
we find
\begin{eqnarray}
\partial_t \hD^r &=& \beta^r \partial_r \hD^r - \hD^r \partial_r \beta^r
+ \frac{1}{a} \partial^2_r \beta^r + \frac{2}{b} \:
\partial_r \left( \frac{\beta^r}{r} \right) \nonumber \\
&+& \frac{\sigma}{3} \left( \frac{1}{a} \partial_r ( \hnabla_m \beta^m ) 
+ 2 \hD^r \hnabla_m \beta^m \right) \nonumber \\
&-& \frac{2}{a} \left( A_a \partial_r \alpha
+ \alpha \partial_r A_a \right) \nonumber \\
&+& 2 \alpha \left( A_a \hD^r - \frac{2}{rb}
\left( A_a - A_b \right) \right) \nonumber \\
&+& \frac{\alpha \xi}{a} \left[ \partial_r A_a
- \frac{2}{3} \: \partial_r K \right.
+ 6 A_a \partial_r \chi \nonumber \\
&+& \left. \left( A_a - A_b \right) \left( \frac{2}{r}
+ \: \frac{\partial_r b}{b} \right) 
- 8 \pi j_r \right] \; ,
\end{eqnarray}
with $j_r$ the (physical) covariant component of the momentum density,
and where as before $\xi$ is an arbitrary parameter such that $\xi >
1/2$, with preferred value $\xi=2$.

Finally, it is also convenient to write the specific form of the
Hamiltonian and momentum constraints. One finds
\begin{eqnarray}
H &=& R - \left( A_a^2 + 2 A_b^2 \right) + \frac{2}{3} \: K 
- 16 \pi \rho = 0 \; , \qquad
\label{eq:sphere-ham} \\
M^r &=& \partial_r A_a - \frac{2}{3} \: \partial_r K
+ 6 A_a \partial_r \chi \nonumber \\
&+& \left( A_a - A_b \right) \left( \frac{2}{r}
+ \frac{\partial_r b}{b} \right)
- 8 \pi j_r = 0 \; .
\label{eq:sphere-mom}
\end{eqnarray}

%%%%%%%%%%%%%%%%%%%%%%%%%%
%%%   REGULARIZATION   %%%
%%%%%%%%%%%%%%%%%%%%%%%%%%

\subsubsection{Regularization}
\label{sec:BSSNregular}

As has already been discussed in~\cite{Alcubierre04a,Ruiz:2007rs},
unless special care is taken, in spherical symmetry the coordinate
singularity at the origin can be a source of serious problems caused
by the lack of regularity of the geometric variables there.  The
problem arises because of the presence of terms in the evolution
equations that go as $1/r$ near the origin.  At the analytic level,
for a regular spacetime one can show that such terms cancel exactly at
the origin, thus ensuring well-behaved solutions.  However, this exact
cancellation usually fails to hold for numerical solutions.  One then
finds that the numerical solution becomes ill-behaved near $r=0$.

There are in fact two different types of regularity conditions that
the geometric variables must satisfy at $r=0$.  The first type of
conditions are simply those imposed by the requirement that the
different variables should have a well defined parity at the origin,
and imply the following behavior for small $r$:
\begin{eqnarray}
\alpha &\sim& \alpha^0 + {\cal O}(r^2) \; , \label{eq:regalpha} \\
\beta^r &\sim& {\cal O}(r) \; , \label{eq:regbeta} \\
a &\sim& a^0 + {\cal O}(r^2) \; , \label{eq:rega} \\
b &\sim& b^0 + {\cal O}(r^2) \; , \label{eq:regb} \\
A_a &\sim& A_a^0 + {\cal O}(r^2) \; , \label{eq:regAa} \\
A_b &\sim& A_b^0 + {\cal O}(r^2) \; , \label{eq:regAb} \\
\hD^r &\sim& {\cal O}(r) \; , \label{eq:regDelta}
\end{eqnarray}
with $\{ \alpha^0,a^0,b^0,A_a^0,A_b^0 \}$ perhaps functions of time,
but not of $r$.

Notice, however, that the above parity conditions are not enough to
guarantee regularity of the system of equations described in the
previous Section. Although most terms involving divisions by powers of
$r$ are indeed manifestly regular given the different parity
conditions (because they involve various derivatives of the geometric
quantities), there are in fact two types of terms that would seem to
remain ill-behaved at the origin.  In particular, the expression for
$\hD^r$ (equation~\eqref{eq:sphere-Delta}) involves the term
$(1-a/b)/r$, while the expressions for the radial component of the
Ricci tensor $R^r_r$ and its trace $R$
(equations~\eqref{eq:sphere-Rrr} and~\eqref{eq:sphere-RSCAL}) have
terms of the form $(1-a/b)/r^2$, which apparently blow up at the
origin. Similarly, in the momentum constraint~\eqref{eq:sphere-mom} we
have a term of the type $(A_a - A_b)/r$, which again would seem to be
ill-behaved at the origin.

The reason why these apparently ill-behaved terms turn out to be
regular after all is a consequence of a second type of regularity
conditions.  These new conditions come from the fact that spacetime
should be locally flat at the origin, and imply that for small $r$ we
must have
\begin{equation}
a - b \sim {\cal O}(r^2) \; , \qquad A_a - A_b \sim {\cal O}(r^2) \; ,
\label{eq:regA-B}
\end{equation}
so that
\begin{equation}
a^0 = b^0 \; , \qquad A_a^0 = A_b^0 \; .
\end{equation}
It turns out that it is not trivial to implement numerically both the
parity regularity conditions and the local flatness regularity
conditions at the same time.  The reason for this is that at $r=0$ we
now have three boundary conditions for just two variables: both the
derivatives of $a$ and $b$ must vanish, plus $a$ and $b$ must be equal
to each other (and the same thing must happen for $A_a$ and $A_b$).

The regularization issue has already been discussed in some detail in
several references~\cite{Arbona98,Alcubierre04a,Ruiz:2007rs}. Here we
will introduce a regularization procedure based on the one presented
in~\cite{Alcubierre04a}, but with one important modification.  We will
then start by introducing an auxiliary variable defined as
\begin{equation}
\lambda := \frac{1}{r^2} \: \left( 1 - \frac{a}{b} \right) \; .
\label{eq:lambda}
\end{equation}
Notice first that in~\cite{Alcubierre04a} the variable $\lambda$ was
in fact defined defined as $\lambda := ( 1 - a/b )/r$. This is just a
small difference of no real consequence. Now, the local-flatness
regularity conditions above imply that close to the origin we must
have
\begin{equation}
\lambda \sim \lambda^0 + {\cal O}(r^2) \; .
\label{eq:regl}
\end{equation}
The main difference with the regularization procedure described
in~\cite{Alcubierre04a} is that we will now also introduce a second
auxiliary variable defined as
\begin{equation}
A_\lambda := \frac{1}{r^2} \: \left( A_a - A_b \right) \; .
\label{eq:Alambda}
\end{equation}
Again, the local--flatness regularity conditions imply that close to
the origin we will have
\begin{equation}
A_\lambda \sim A_\lambda^0 + {\cal O}(r^2) \; .
\label{eq:regAl}
\end{equation}

Having introduced $\lambda$ and $A_\lambda$ we can rewrite all
apparently ill-behaved terms in the BSSN equations in terms of these
quantities so that the equations now look regular.  In particular, the
expression for $\hD^r$ becomes
\begin{equation}
\hD^r = \frac{1}{a} \left[ \frac{\partial_r a}{2a}
- \frac{\partial_r b}{b} - 2 r \lambda \right] \; ,
\label{eq:sphere-Delta-reg}
\end{equation}
while $R^r_r$ and $R$ take the form
\begin{eqnarray}
R^r_r &=& - \frac{1}{a e^{4 \chi}} \left[ \frac{\partial^2_r a}{2a} 
- a \partial_r \hD^r  - \frac{3}{4} \left( \frac{\partial_r a}{a} \right)^2 \right.
\nonumber \\
&+& \frac{1}{2} \left( \frac{\partial_r b}{b} \right)^2
- \frac{1}{2} \hD^r \partial_r a + \frac{\partial_r a}{rb}
+ 2 \lambda \left( 1 + \frac{r \partial_r b}{b} \right)
\nonumber \\
&+& 4 \left. \partial^2_r \chi - 2 \partial_r \chi \left( \frac{\partial_r a}{a}
- \frac{\partial_r b}{b} - \frac{2}{r} \right) \right] ,
\label{eq:sphere-Rrr-reg} \\
R &=& - \frac{1}{a e^{4 \chi}} \left[ \frac{\partial^2_r a}{2a}
+ \frac{\partial^2_r b}{b} - a \partial_r \hD^r
- \left( \frac{\partial_r a}{a} \right)^2
\right. \nonumber \\
&+& \frac{1}{2} \left( \frac{\partial_r b}{b} \right)^2 + \frac{2}{rb}
\left( 3 - \frac{a}{b} \right) \partial_r b \nonumber \\
&+& 4 \lambda + 8 \left( \partial^2_r \chi + ( \partial_r \chi )^2 \right)
\nonumber \\
&-& \left. 8 \partial_r \chi \left( \frac{\partial_r a}{2a}
- \frac{\partial_r b}{b} - \frac{2}{r} \right) \right] .
\label{eq:sphere-RSCAL-reg}
\end{eqnarray}
Similarly, the momentum constraint now becomes
\begin{eqnarray}
M^r &=& \partial_r A_a - \frac{2}{3} \: \partial_r K
+ 6 A_a \partial_r \chi \nonumber \\
&+& A_\lambda \left( 2 r + r^2 \: \frac{\partial_r b}{b} \right)
- 8 \pi j_r = 0 \; .
\label{eq:sphere-mom-reg}
\end{eqnarray}

Of course, at this point we haven't really fixed the problem, all we
have actually done is to define some new variables as short-hands where
we have hidden the ill-behaved terms.  The way to solve the
problem is to promote $\lambda$ and $A_\lambda$ to independent
variables (with initial data given through their definitions), and
find evolution equations for them that are manifestly regular.

The evolution equation for $\lambda$ can be found directly from its
definition and the evolution equations for $a$ and $b$
(equations~\eqref{eq:sphere-adot} and~\eqref{eq:sphere-bdot}), and
turns out to be
\begin{equation}
\partial_t \lambda = \beta^r \partial_r \lambda + \frac{2}{r}
\left[ \beta^r \lambda - \frac{a}{b} \: \partial_r \left( \frac{\beta^r}{r}
\right) \right] + \frac{2 \alpha a}{b} \: A_\lambda \; .
\label{eq:sphere-lambdadot}
\end{equation}
Notice that this equation is manifestly regular as long as
\mbox{$\beta^r \sim {\cal O}(r)$} for small $r$, and $A_\lambda$ itself
remains regular.  Notice also that the equation does not involve
$\sigma$, so it has the same form for Eulerian or Lagrangian
evolutions.

In order to find the evolution equation for $A_\lambda$ it is
important to notice first that the traceless condition \mbox{$A_a + 2
  A_b = 0$} implies that $A_\lambda$ and $A_a$ are in fact related
through
\begin{equation}
A_\lambda = \frac{3 A_a}{2 r^2} \; .
\label{eq:Alambda2}
\end{equation}
This actually implies that for $A_\lambda$ to remain regular we must
ask for \mbox{$A_a \sim {\cal O}(r^2)$} near the origin (or in other
words: \mbox{$A^0_a = A^0_b = 0$}).

We can now use the evolution equation for $A_a$,
equation~\eqref{eq:sphere-Adot}, to obtain the evolution equation for
$A_\lambda$.  After a long algebra, in which one needs to take care of
keeping together several terms that might be ill-behaved individually
(particularly terms of the form $\partial_r (F/r)$, with $F$ some
function that behaves as $F \sim {\cal O}(r)$ near the origin), one
finds
\begin{eqnarray}
\partial_t A_\lambda &=& \beta^r \partial_r A_\lambda
+ 2 A_\lambda \frac{\beta^r}{r} \nonumber \\
&& \hspace{-10mm} - \frac{1}{r a e^{4 \chi}}
\left[ \partial_r \left( \frac{\partial_r \alpha}{r} \right)
- \frac{\partial_r \alpha}{2 r}
\left( \frac{\partial_r a}{a} +  \frac{\partial_r b}{b}
+ 8 \partial_r \chi \right) \right] \nonumber \\
&& \hspace{-10mm} - \frac{\alpha} {r a e^{4 \chi}}
\left[ 2 \partial_r \left( \frac{\partial_r \chi}{r} \right) 
- \frac{\partial_r \chi}{r}
\left( \frac{\partial_r a}{a} +  \frac{\partial_r b}{b}
+ 4 \partial_r \chi \right) \right] \nonumber \\
&& \hspace{-10mm} + \frac{\alpha} {a e^{4 \chi}}
\left[ \frac{b}{2a} \: \partial^2_r \lambda
+ \frac{a}{r} \: \partial_r \left( \frac{\hD^r}{r} \right) \right.
\nonumber \\
&& \hspace{-10mm} + \frac{\partial_r \lambda}{r} \left( 1 + \frac{2b}{a}
- \frac{r b}{2} \: \hD^r \right)
+ \frac{\partial_r a}{a r^2} \left( \frac{3}{4} \frac{\partial_r a}{a}
- \frac{\partial_r b}{b} \right) \nonumber \\
&& \hspace{-10mm} \left. - \frac{\lambda}{r} \left( b \hD^r
+ 2 \: \frac{\partial_r b}{b} \right) + \frac{b}{a} \: \lambda^2 \right]
\nonumber \\
&& \hspace{-10mm} + \alpha K A_\lambda - 8 \pi \alpha S_\lambda \; ,
\label{eq:sphere-Alambdadot}
\end{eqnarray}
where $S_\lambda := (S_a - S_b)/r^2$.  The above equation is now manifestly
regular in all its terms.

Notice that at this point one can in fact just choose to ignore the
evolution equation for $A_a$, evolve $A_\lambda$
using~\eqref{eq:sphere-Alambdadot}, and later recover $A_a$ though the
relation $A_a=2 r^2 A_\lambda/3$.  This is in fact what we do in the
numerical code used for the numerical examples of the next Section,
since in practice it seems to reduce considerably the size of the
numerical error (though evolving both $A_a$ and $A_\lambda$
independently also results in regular and stable evolutions).

%%%%%%%%%%%%%%%%%%%%%%%%%%%%%%
%%%   NUMERICAL EXAMPLES   %%%
%%%%%%%%%%%%%%%%%%%%%%%%%%%%%%

\section{Some numerical examples}
\label{sec:examples}

We have constructed a numerical code using the regularized BSSN
formulation in spherical symmetry described in the last Section.  The
code uses a method of lines algorithm, with either second order 3-step
iterative Crank-Nicholson (ICN) or fourth order Runge--Kutta (RK4) as
the time integrator, and second or fourth order centered differences
in space.  This code turns out to be very robust, stable, and well
behaved at the origin.

Here we will present some examples of numerical simulations using this
code that involve both pure gauge dynamics in vacuum situations, and
simulations with a non-zero matter field.  We will also consider
simulations of a Schwarzschild black hole, which is a special case
since it is in fact not regular at the origin.

%%%%%%%%%%%%%%%%%%%%
%%%   HARMONIC   %%%
%%%%%%%%%%%%%%%%%%%%

\subsection{Pure gauge dynamics}

As a first example we will consider a pure gauge pulse propagating
through the numerical spacetime.  We consider initial data
corresponding to Minkowski spacetime:
\begin{eqnarray}
&& \chi = 0 \; , \\
&& a = b = 1 \; , \\
&& A_a = A_b = K = 0 \; , \\
&& \hD^r = 0 \; ,
\end{eqnarray}
which also imply $\lambda = A_\lambda=0$.  Non-trivial gauge
dynamics are obtained by choosing an initial lapse
with a Gaussian profile of the form
\begin{equation}
\alpha = 1 + \frac{\alpha_0 r^2}{1+r^2}
\left[ e^{-(r-r_0)^2/\sigma^2}
+ e^{-(r+r_0)^2/\sigma^2} \right] \; ,
\end{equation}
with $\alpha_0$ some initial amplitude, $r_0$ the center of the
Gaussian and $\sigma$ its width. Notice that we have multiplied the
whole expression with $r^2/(1+r^2)$ and have in fact added two
symmetric Gaussians (centered at $r=r_0$ and $r=-r0$). This is done in
order to make sure that the initial lapse is both an even function of
$r$, and vanishes at $r=0$.  Having set up this initial lapse we
evolve the system using harmonic slicing (with zero shift):
\begin{equation}
\partial_t \alpha = - \alpha^2 K \; .
\end{equation}

\begin{figure}
\epsfxsize=110mm
\centerline{\epsfbox{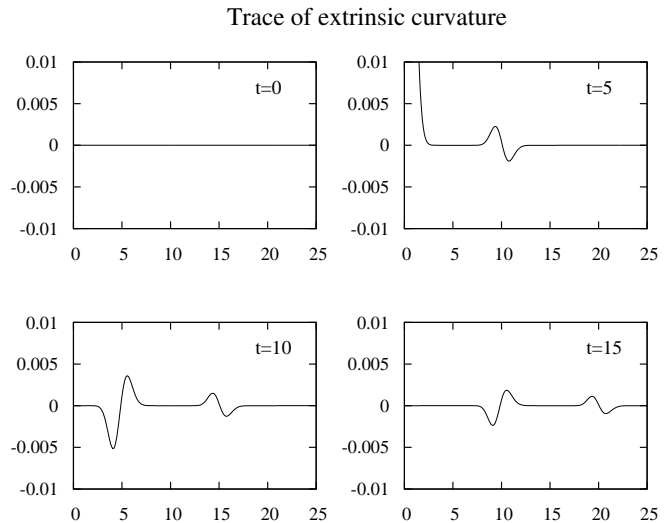}}
\caption{Evolution of the trace of the extrinsic curvature $K$.  The
  different panels correspond to times $t=0,5,10,15$.}
\label{fig:trK_harmonic}
\end{figure}

For the simulation shown below we have chosen the initial data
parameters as $\alpha_0=0.01$, $r_0=5$, $\sigma=1$, with grid
parameters given by $\Delta r = 0.1$ and $\Delta t = \Delta r / 2$.
During the simulation the initial pulse first separates in two smaller
pulses propagating in opposite directions (due to the time-symmetry of
the initial data). The inward moving pulse later implodes through the
origin at $t \sim 5$ and starts moving outward.
Figure~\ref{fig:trK_harmonic} shows a snapshot of the evolution of the
extrinsic curvature $K$ at times $t=0,5,10,15$.  Notice that at $t=5$
the value of $K$ at the origin is too large for the chosen scale (it
reaches a value of $\sim 0.1$), however the evolution always remains
well-behaved and the value of $K$ at the origin later returns to zero
as the pulse moves outward.

\begin{figure}
\epsfxsize=110mm
\centerline{\epsfbox{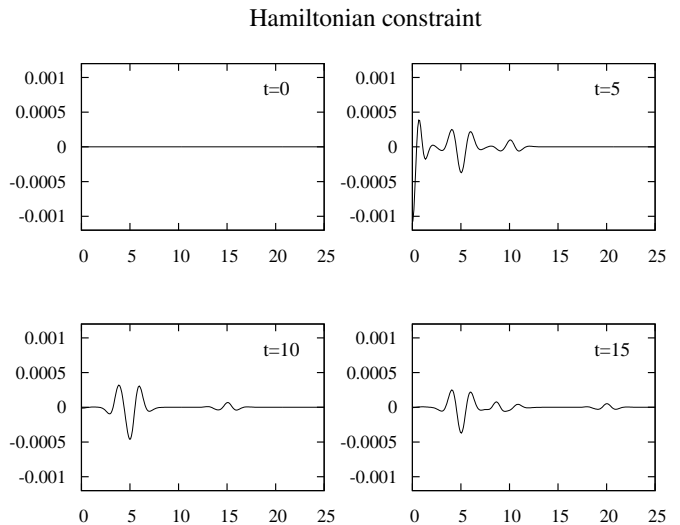}}
\caption{Evolution of the Hamiltonian constraint. The different panels
  correspond to times $t=0,5,10,15$.}
\label{fig:ham_harmonic}
\end{figure}

Figure~\ref{fig:ham_harmonic} shows the evolution of the Hamiltonian
constraint for this simulation.  Notice again that it remains
well-behaved as the pulse goes through the origin.  A small remnant of
constraint violation that does not propagate away can also be clearly
seen centered at the initial position of the pulse $r \sim 5$.

\begin{figure}
\epsfxsize=90mm
\centerline{\epsfbox{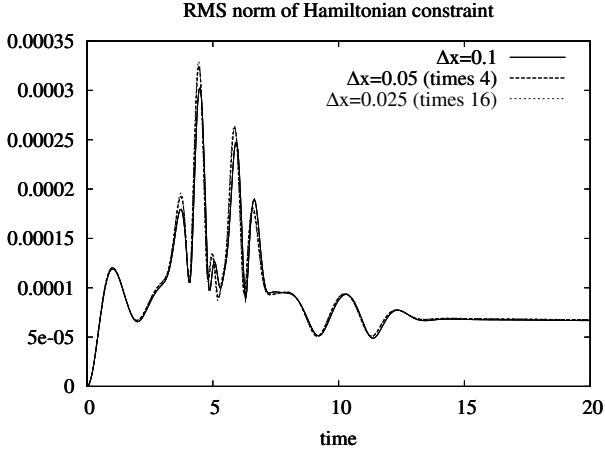}}
\caption{RMS norm of the Hamiltonian constraint as a function of time,
  for simulations at three different resolutions: $\Delta r = 0.1$,
  $\Delta r = 0.05$ and $\Delta r = 0.025$. The two highest
  resolutions have been rescaled by factors of 4 and 16 respectively.}
\label{fig:conv_harmonic}
\end{figure}

Finally, we will consider the issue of convergence of the code.
Figure~\ref{fig:conv_harmonic} shows the root-mean-square (RMS) norm
of the Hamiltonian constraint as a function of time, for simulations
at three different resolutions, $\Delta r = 0.1$, $\Delta r = 0.05$
and $\Delta r = 0.025$ (keeping always the same ratio $\Delta t /
\Delta r = 0.5$).  For the two highest resolutions, the norms have
been rescaled by the corresponding factors expected for second order
convergence, 4 and 16 respectively.  The fact that all three lines
coincide indicates that the code is indeed converging to second order.

%%%%%%%%%%%%%%%%%%%%%%%%
%%%   SCALAR FIELD   %%%
%%%%%%%%%%%%%%%%%%%%%%%%

\subsection{Scalar field}

The second set of simulations are somewhat more interesting since they
evolve a non-zero matter field and therefore contain true dynamics.
The matter field will correspond to a real scalar field $\Phi$ with
stress-energy tensor given by:
\begin{equation}
T_{\mu \nu} = \partial_\mu \Phi \partial_\nu \Phi
- \frac{1}{2} \: g_{\mu \nu} \: \partial^\alpha \Phi \partial_\alpha \Phi \; .
\end{equation}

Let us now consider the case of spherical symmetry, and define the
auxiliary first order quantities:
\begin{eqnarray}
\Pi &:=& n^\mu \partial_\mu \Phi = \frac{1}{\alpha} \left(
\partial_t \Phi - \beta^r \partial_r \Phi \right) \; , \\
\Psi &:=& \partial_r \Phi \; .
\end{eqnarray}
The corresponding energy density $\rho$, momentum density $j^r$, and
stress tensor $S_{ij}$ that appear in the 3+1 equations are then given
by
\begin{eqnarray}
\rho &:=& n^\mu n^\nu T_{\mu \nu} = \frac{1}{2} \left( \Pi^2
+ \frac{\Psi^2}{a e^{4 \chi}} \right) \; , \\
j^r &:=& - P^{r \mu} n^{\nu} T_{\mu \nu} =  - \Pi \Psi \; , \\
S_a &:=& T^r_r = \frac{1}{2} \left( \Pi^2
+ \frac{\Psi^2}{a e^{4 \chi}} \right) \; , \\
S_b &:=& T^\theta_\theta = \frac{1}{2} \left( \Pi^2
- \frac{\Psi^2}{a e^{4 \chi}} \right) \; .
\end{eqnarray}

The scalar field evolves through the simple wave equation \mbox{$\Box
  \Phi = 0$}, which in this case can be written as the following
system of first order equations:
\begin{eqnarray}
\partial_t \Phi &=& \beta^r \partial_r \Phi  + \alpha \Pi \; , \\
\partial_t \Psi &=& \beta^r \partial_r \Psi + \Psi \partial_r \beta^r
+ \partial_r \left( \alpha \Pi \right) \; , \\
\partial_t \Pi &=& \beta^r \partial_r \Pi + \frac{\alpha}{a  e^{4 \chi}}
\left[ \partial_r \Psi \right. \nonumber \\
&+& \left. \Psi \left( \frac{2}{r} - \frac{\partial_r a}{2a}
+ \frac{\partial_r b}{b} + 2 \partial_r \chi \right) \right] \nonumber \\
&+& \frac{\Psi}{a  e^{4 \chi}} \: \partial_r \alpha + \alpha K \Pi \; .
\end{eqnarray}

For the initial data, we again choose an initial Gaussian profile for
the scalar field of the form
\begin{equation}
\Phi = \frac{\Phi_0 r^2}{1+r^2} \left[
  e^{-(r-r_0)^2/\sigma^2} + e^{-(r+r_0)^2/\sigma^2} \right] \; ,
\end{equation}
with $\Phi_0$ the initial amplitude, $r_0$ the center of the Gaussian
and $\sigma$ its width.  We also assume time symmetry, so that at
$t=0$ we have $K_{ij}=0$ and $\Pi=0$, which implies that the momentum
constraint is identically satisfied.

Finally, we choose a conformally flat metric with \mbox{$a=b=1$}, and
solve for the conformal factor $\psi=e^\chi$ using the Hamiltonian
constraint, which in this case takes the form
\begin{equation}
\partial_t^2 \psi + \frac{2}{r} \: \partial_r \psi + 2 \pi \psi^5 \rho = 0 \; .
\end{equation}
Notice that when we substitute the value of $\rho$ given above (with
$\Pi=0$), this equation reduces to
\begin{equation}
\partial_t^2 \psi + \frac{2}{r} \: \partial_r \psi
+ \left( \pi \Psi^2 \right) \psi = 0 \; .
\end{equation}
This is clearly a linear equation for $\psi$, which simplifies
considerably its numerical solution (we can simply use centered
spatial differences and invert the resulting tridiagonal matrix
directly).

As gauge conditions we have chosen zero shift and 1+log slicing, which
is given by
\begin{equation}
\partial_t \alpha = - 2 \alpha K \; .
\end{equation}

For the specific simulation discussed here, we have taken as initial
data parameters: $\Phi_0=0.04$, $r_0=5$, $\sigma=1$.  We have chosen
this initial data to be rather strong, but not quite strong enough to
collapse to a black hole. This is on purpose since we are interested
in the regularity at the origin, and the collapse of the lapse
associated with the formation of a black hole makes this issue less
relevant as everything just freezes close to the origin.  In
Section~\ref{sec:schwarz} below we will consider the case of a single
Schwarzschild black hole.

\begin{figure}
\epsfxsize=110mm
\centerline{\epsfbox{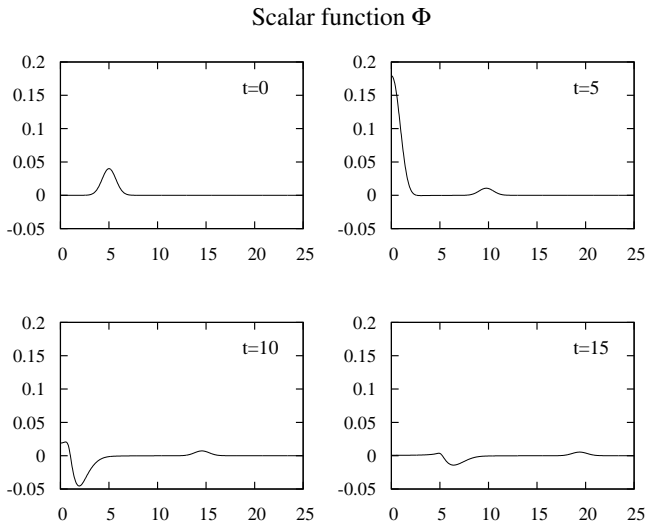}}
\caption{Evolution of the scalar field $\Phi$.  The
  different panels correspond to times $t=0,5,10,15$.}
\label{fig:phi_scalar}
\end{figure}

In Figure~\ref{fig:phi_scalar} we show snapshots of the evolution of
the scalar field $\Phi$, for a numerical simulation using a grid
spacing \mbox{$\Delta r = 0.025$} and time step $\Delta t = \Delta r /
2$.  One can clearly see how the initial pulse separates into ingoing
and outgoing pieces.  The ingoing part then implodes through the
origin and starts moving out, though significantly deformed.

\begin{figure}
\epsfxsize=90mm
\centerline{\epsfbox{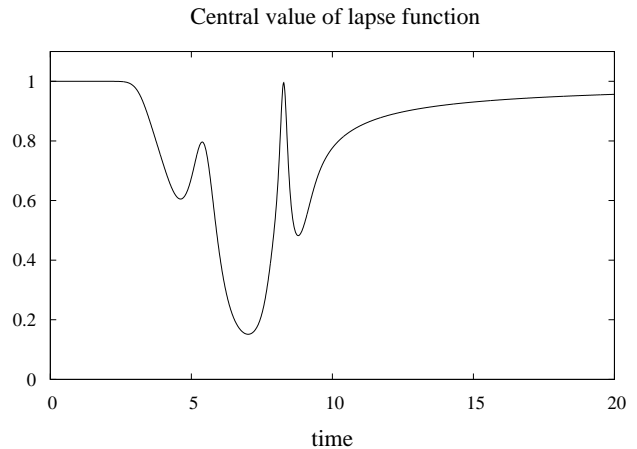}}
\caption{Central value of the lapse as a function of time.}
\label{fig:alpha_scalar}
\end{figure}

Figure~\ref{fig:alpha_scalar} shows the evolution of the central value
of the lapse function $\alpha$ as a function of time (actually the
value at $r=\Delta r /2$ since the origin itself is staggered).
Notice how at $t\sim 7$ the central value of the lapse drops below
$0.2$, indicating a very strong gravitational field.  However, the
lapse later bounces and returns towards 1, and no black hole forms.
The simulation remains well behaved throughout, and the origin remains
regular.

\begin{figure}
\epsfxsize=90mm
\centerline{\epsfbox{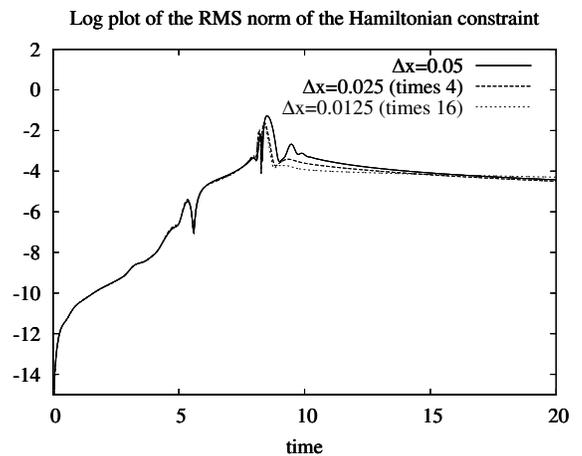}}
\caption{Logarithm of the RMS norm of the Hamiltonian constraint as a
  function of time, for simulations at three different resolutions:
  $\Delta r = 0.05$, $\Delta r = 0.025$ and $\Delta r = 0.0125$. The
  two highest resolutions have been rescaled by factors of 4 and 16
  respectively.}
\label{fig:conv_scalar}
\end{figure}

Finally, in Figure~\ref{fig:conv_scalar} we show again a plot of the
RMS norm of the Hamiltonian constraint for three different
resolutions, $\Delta r = 0.05$, $\Delta r = 0.025$ and $\Delta r =
0.0125$ (due to the large dynamical range, the plot is now
logarithmic).  Again, the higher resolution runs have been rescaled by
factors of 4 and 16 respectively.  The fact that the lines lie on top
of each other shows that the code is converging to second order, as
expected. One can notice that from \mbox{$t \sim 7$} to \mbox{$t \sim
  15$} the convergence is less than perfect and the lines do not align
precisely.  This behaviour is a reflection of the fact that the
gravitational field is very strong so that very high resolution is
needed to adequately capture the situation.

%%%%%%%%%%%%%%%%%%%%%%%%%
%%%   SCHWARZSCHILD   %%%
%%%%%%%%%%%%%%%%%%%%%%%%%

\subsection{Schwarzschild black hole}
\label{sec:schwarz}

As our final example we will choose a Schwarzschild black hole, so
again we are back in vacuum.  The Schwarzschild solution is static in
standard coordinates, but these coordinates are ill-behaved at the
horizon.  We will therefore use isotropic coordinates in which the
initial spatial metric takes the form
\begin{equation}
dl^2 = \psi^4 \left( dr^2 + r^2 d \Omega^2 \right) \; ,
\end{equation}
where the conformal factor $\psi$ is given by
\begin{equation}
\psi = 1 + M/2r \; ,
\end{equation}
and with $M$ the mass of the black hole.  As is well known, the
Schwarzschild solution in isotropic coordinates has the topology of a
wormhole (Einstein--Rosen bridge), with the throat located at $r=M/2$
(coincident with the horizon at $t=0$), and with a coordinate
singularity at $r=0$ which corresponds to the compactification of the
asymptotic flat region on the other side of the wormhole.

The presence of the coordinate singularity at \mbox{$r=0$} implies
that our regularization procedure is now wrong since it was based on
the idea of space being locally flat at the origin, which is clearly
not the case here.  The origin is not even part of space, which is why
this type of initial data is known as puncture initial data (space in
fact corresponds to $\Re^3$ minus the point at the origin, so it is a
``punctured'' $\Re^3$).  The main consequence of this is that using
the regularization procedure now fails and the simulations quickly
crash.  However, we have found that if we simply turn off the
regularization, and run without introducing the variables $\lambda$
and $A_\lambda$, we can have very stable and accurate simulations with
the exception of the first few grid points closer to the origin where
the code fails to converge (but we do find convergence away from these
points as the plots below show).  A more careful look at the data
shows that, for the simulations described below, close to the origin
we have \mbox{$A_a \sim r$}, so that \mbox{$A_\lambda \sim 1/r$}
(confront equation~\eqref{eq:Alambda2}), which explains why the
regularization procedure fails. A more detailed analysis of the
behaviour at the origin for black hole simulations is clearly needed,
but this is outside the scope of this paper.

For all the simulations shown here we have chosen maximal slicing.
This corresponds to the condition \mbox{$K=\partial_t K=0$}, which
leads to the following equation for the lapse function:
\begin{eqnarray}
\partial_r^2 \alpha + \left( \frac{2}{r} -
\frac{\partial_r a}{2a} + \frac{\partial_r b}{b} + 2 \partial_r \chi
\right) \partial_r \alpha \hspace{8mm} && \nonumber \\
- \alpha a e^{4 \chi} \left[ K_{ij} K^{ij}
+ 4 \pi \left( \rho + S_a + 2 S_b \right) \right] &=& 0 ,
\hspace{10mm}
\end{eqnarray}
where $K_{ij} K^{ij} = A_a^2 + 2 A_b^2 + K^2/3$.  Notice that this is
again a linear equation for $\alpha$, which can be solved by direct
matrix inversion.  The reason for choosing maximal slicing is to make
sure that the lapse collapses at \mbox{$r=0$}, which would not happen
with 1+log slicing.  This is because the origin is in fact an infinite
proper distance away, so that any slicing condition with a finite
speed of propagation would never change the value of the lapse
there.~\footnote{One could still use 1+log slicing if one chooses a
  pre-collapsed initial lapse.  This is in fact what is typically done
  in 3D black hole simulations.}

For the shift we have chosen a Gamma driver condition of the form
discussed in Section~\ref{sec:gammadriver}.  In the particular
case of spherical symmetry this condition reduces to:
\begin{equation}
\partial^2_t \beta^r = \frac{3}{4} \: \partial_t \hD^r
- \eta \: \partial_t \beta^i \; .
\end{equation}
Here we have already chosen the coefficient of the term
\mbox{$\partial_t \hD^r$} equal to $3/4$ in order to have an
asymptotic gauge speed equal to $1$.  The condition above is solved in
first order form by introducing the time derivative of the shift as an
auxiliary quantity, so that we in fact solve the system:
\begin{eqnarray}
\partial_t \beta^r &=& B^r \; , \\
\partial_t B^r &=& \frac{3}{4} \: \partial_t \hD^r - \eta B^r \; .
\end{eqnarray}
Notice that this is the same shift condition (with minor variations)
that is currently being used in most 3D codes that evolve black hole
spacetimes.  In the simulations shown below, the damping coefficient
is always taken to be $\eta = 2$.

We still need to mention one final ingredient that goes into these
simulations. Following~\cite{Campanelli:2005dd,Campanelli:2006gf}, we
have found that the simulations are better behaved if instead of
evolving the singular conformal factor $\chi$ directly, we evolve the
quantity \mbox{$X := e^{-2 \chi}$}.~\footnote{Notice that
  in~\cite{Campanelli:2005dd,Campanelli:2006gf} they in fact evolve
  the quantity \mbox{$\chi := e^{-4 \phi}$}. Here our notation is
  different, so that $\chi$ plays the role of $\phi$, and $X$ the role
  of $\chi$, we also use a second power instead of fourth power since
  we find this to work better in our case.}

We are now ready to describe the numerical simulations. In all our
simulations we have chosen the mass of the black hole to be $M=1$.  We
have chosen a grid spacing of $\Delta r =0.01$ and time step of
$\Delta t = 0.005$.  We have also used 10,000 grid points in order to
place the boundaries sufficiently far away so as not to have large
errors from the boundaries affect the evolution.~\footnote{The boundary
  conditions chosen are stable and well behaved, but we will not
  discuss them in any detail here. We are preparing a paper where we
  will concentrate on the boundary conditions.}

\begin{figure}[ht]
\epsfxsize=110mm
\centerline{\epsfbox{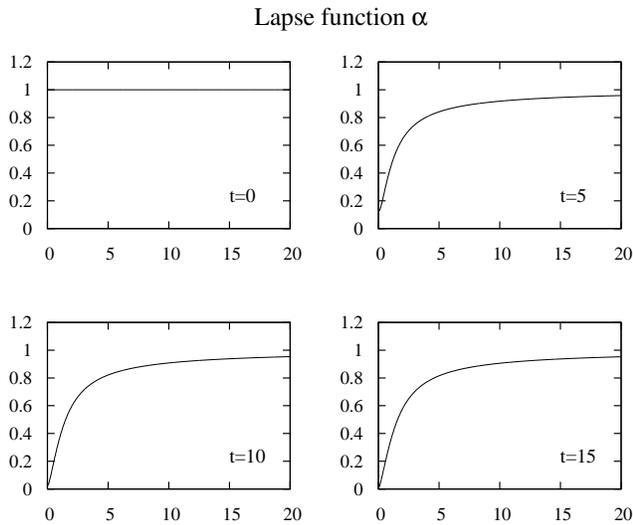}}
\caption{Evolution of the lapse function $\alpha$.  The
  different panels correspond to times $t=0,5,10,15$.}
\label{fig:alpha_schwarz}
\end{figure}

\begin{figure}[ht]
\epsfxsize=110mm
\centerline{\epsfbox{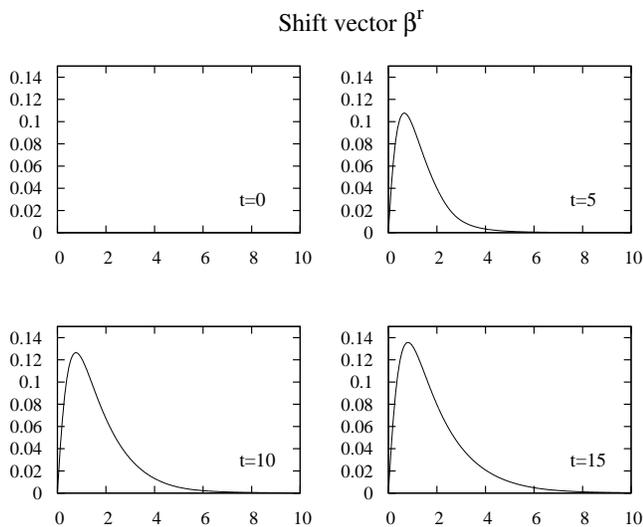}}
\caption{Evolution of the radial component of the shift vector
  $\beta^r$.  The different panels correspond to times $t=0,5,10,15$.}
\label{fig:beta_schwarz}
\end{figure}

\begin{figure}[ht]
\epsfxsize=110mm
\centerline{\epsfbox{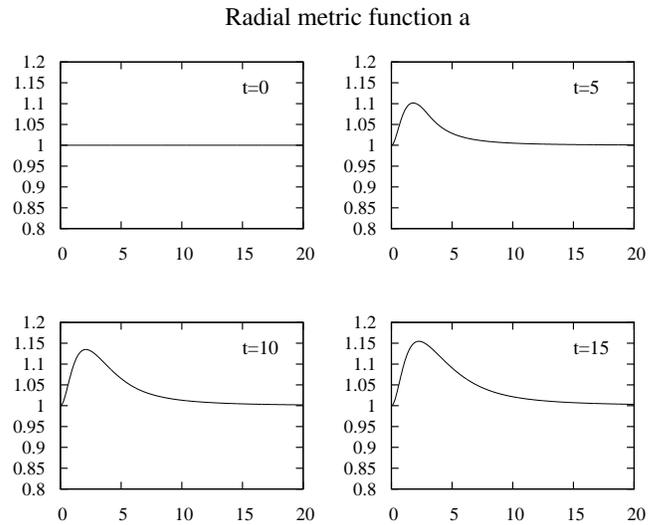}}
\caption{Evolution of the radial metric function $a$.  The different
  panels correspond to times $t=0,5,10,15$.}
\label{fig:A_schwarz}
\end{figure}

\begin{figure}[ht]
\epsfxsize=110mm
\centerline{\epsfbox{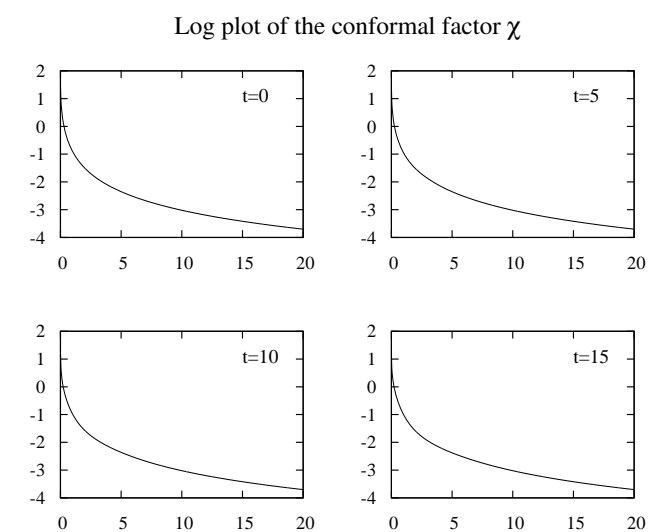}}
\caption{Evolution of the log of the conformal factor $\chi$.  The
  different panels correspond to times $t=0,5,10,15$.}
\label{fig:phi_schwarz}
\end{figure}

Figures~\ref{fig:alpha_schwarz}-\ref{fig:phi_schwarz} show snapshots
of the evolution of the lapse function $\alpha$, the radial component
of the shift vector $\beta^r$, the radial metric component $a$ and the
conformal factor $\chi$ at times $t=0,5,10,15$.  Do notice that in
order to better appreciate the plots, for the conformal factor $\chi$
we have used a log plot, while for the shift we plot a smaller radial
domain.

The first thing to notice from these plots is the fact that the
simulation is well behaved.  The lapse collapses to $0$ at the origin
(but with a non-zero derivative there as expected from the gauge
conditions used, see {\em e.g.}~\cite{Hannam:2006vv}).  The shift
grows with time to counteract the slice stretching effect, but becomes
almost stationary at late times.  While the conformal factor, though
still singular at the origin also remains well behaved.

\begin{figure}[!t]
\epsfxsize=100mm
\centerline{\epsfbox{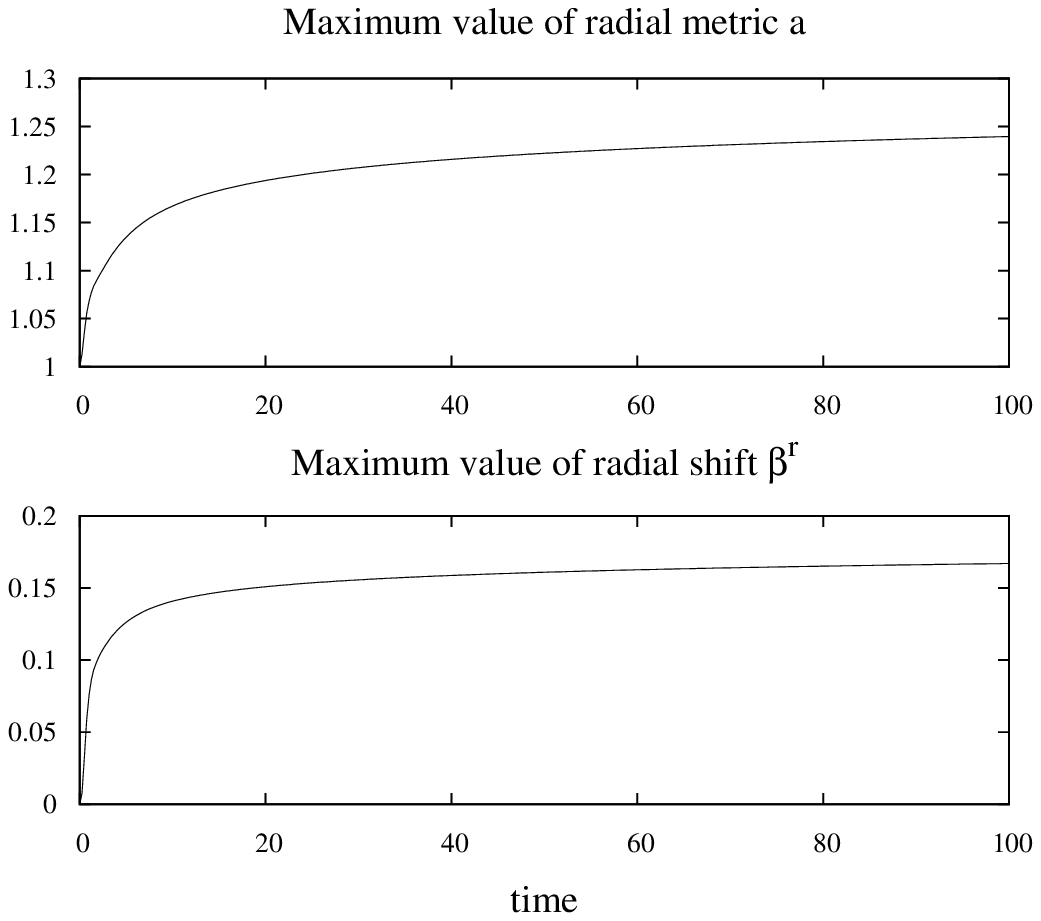}}
\caption{Evolution of the maximum value of the radial metric component
  $a$ (upper panel), and the radial component of the shift vector
  $\beta^r$ (lower panel).}
\label{fig:max_schwarz}
\end{figure}

In order to show that the simulations remain well behaved for long
times, and in fact reach an almost stationary state, in
Figure~\ref{fig:max_schwarz} below we show the evolution of the
maximum value of the radial metric component $a$ and the radial shift
vector $\beta^r$ up to $t=100$.  In both plots we can see that
initially the maximum values grow rapidly, but this behaviour is later
replaced by a very slow upward drift (this drift is well known from 3D
simulations and is a consequence of the gauge conditions, particularly
the damping term in the Gamma driver shift condition).

\begin{figure}[!t]
\epsfxsize=100mm
\centerline{\epsfbox{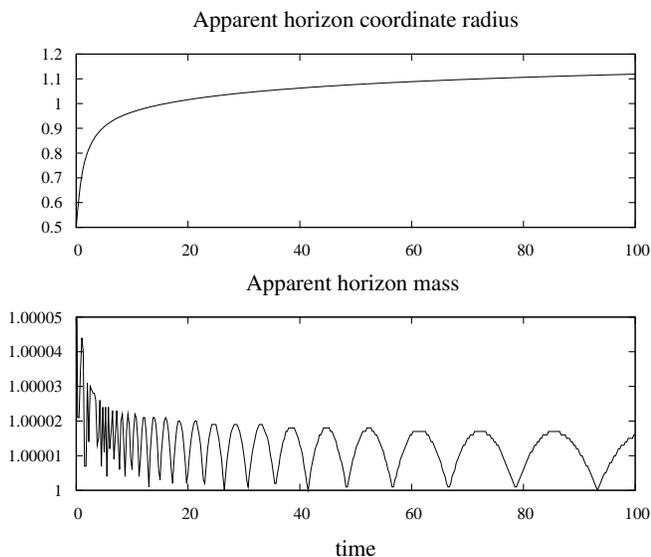}}
\caption{Coordinate position of the apparent horizon (upper panel),
  and apparent horizon mass $M_{ah} = (A_ah/16 \pi)^{1/2}$ (lower
  panel), as functions of time.}
\label{fig:ah_schwarz}
\end{figure}

Figure~\ref{fig:ah_schwarz} shows the time evolution of the coordinate
position of the apparent horizon and the apparent horizon mass
(defined in terms of its area $A_ah$ as $M_{ah} = (A_ah/16
\pi)^{1/2}$).  One can notice how the radial position of the horizon
drifts outward from $r=0.5$ initially to $r \sim 1.1$ at the end of
the simulation.  On the other hand, the horizon mass remains within
$0.005 \%$ of unity throughout the entire simulation.

This shows that the spherically symmetric BSSN code with maximal
slicing and a Gamma driver shift condition can successfully and
accurately simulate a black hole spacetime for a very long time
without the need to excise the black hole interior.

%%%%%%%%%%%%%%%%%%%%%%%
%%%   CONCLUSIONS   %%%
%%%%%%%%%%%%%%%%%%%%%%%

\section{Conclusions}
\label{sec:conclusions}

Following Brown~\cite{Brown:2009dd}, in this paper we have described
how to modify standard hyperbolic formulations of the 3+1 evolution
equations of General Relativity in such a way that all auxiliary
quantities are true tensors, thus allowing for these formulations to
be used with curvilinear sets of coordinates.  We have considered in
particular both the Nagy-Ortiz-Reula (NOR) and the
Baumgarte-Shapiro-Shibata-Nakamura (BSSN) formulations, but the main
ideas presented here can in principle be applied in general.

The key idea has been that instead of using contracted Christoffel
symbols as auxiliary variables, one should use the difference between
the physical Christoffel symbols and the Christoffel symbols
associated with a background flat metric in the curvilinear
coordinates, since the difference of Christoffel symbols always
behaves as a true tensor.

Also, in the particular case of the BSSN formulation, one should not
force the conformal volume element (the metric determinant) to be
equal to 1, but rather to be equal to its initial value in the
curvilinear coordinates. One important consequence of this choice is
that all dynamical geometric quantities now remain true tensors
instead of tensor densities as in standard BSSN.  One is also free to
choose how this conformal volume element will evolve in time.  Two
``natural'' choices present themselves: the Lagrangian approach where
the conformal volume elements remain constant along time lines, and
the Eulerian approach where they remain constant along the normal
direction to the hypersurfaces.  Standard BSSN can then be shown to be
equivalent to the Lagrangian approach.

Having developed the general formalism, we considered as an example
the particular case of BSSN in spherical symmetry, and studied in some
detail the important problem of the regularity of the equations at the
origin.  For this we introduced extra auxiliary quantities that
allowed us to impose the ``local flatness'' regularity condition in a
consistent way.  Our regularization algorithm is similar to the one
presented in~\cite{Alcubierre04a}, but it has been modified in a way
that makes it more general and easier to implement.  It is important
to mention that this regularization assumes that spacetime is regular
at the origin and as such does not work for the case of black hole
spacetimes where the origin is in fact a compactification of an
asymtptic infinity on the other side of the Einstein-Rosen bridge.

Finally, we presented a series of numerical simulations of our BSSN
code in spherical symmetry, and we showed that the code was capable of
evolving both regular spacetimes (with and without matter), as well as
black hole spacetimes in a stable, accurate and robust way.

%%%%%%%%%%%%%%%%%%%%%%%%%%%
%%%   ACKNOWLEDGMENTS   %%%
%%%%%%%%%%%%%%%%%%%%%%%%%%%

\begin{acknowledgments}

The authors wish to thank Dario N\'u\~nez for many useful discussions and
comments.

This work was supported in part by Direcci\'on General de Estudios de
Posgrado (DGEP-UNAM), by CONACyT through grant 82787, and by
DGAPA-UNAM through grants IN113907 and IN115310. M.D.M. also
acknowledges a CONACyT postgraduate scholarship.

\end{acknowledgments}

%%%%%%%%%%%%%%%%%%%%%%
%%%   REFERENCES   %%%
%%%%%%%%%%%%%%%%%%%%%%

\bibliographystyle{bibtex/prsty}
\bibliography{bibtex/referencias}

%%%%%%%%%%%%%%%
%%%   END   %%%
%%%%%%%%%%%%%%%

\end{document}